\documentclass[a4paper,10pt,twocolumn]{article}
\usepackage{cite}
\usepackage{amsmath,amssymb,amsfonts}
\usepackage{algorithm,algpseudocode}
\usepackage{graphicx}
\usepackage{textcomp}
\usepackage{multirow}

\usepackage[
  top=20mm,
  bottom=20mm,
  left=18mm,
  right=18mm
]{geometry}

\usepackage{bm}

\newenvironment{keywords}{\par\noindent\textbf{Keywords:} }{\par}
\newdimen\titlepgskip

\title{Comparison of Algebraic Block Multi-Coloring and Leiden Methods for Parallel Preconditioning in the ICCG Method
\thanks{This work was supported by the Japan Society for the Promotion of Science KAKENHI under grant number 25K15136.}
}

\author{Tomohiro Suzuki \\
 Department of Interdisciplinary Research, Division of Engineering \\
 University of Yamanashi \\
 Takeda 4-3-11, Kofu, Yamanashi 400-8511 Japan \\
 \texttt{stomo@yamanashi.ac.jp}
}
\date{}

\begin{document}
\maketitle

% \markboth
% {Author \headeretal: Preparation of Papers for IEEE TRANSACTIONS and JOURNALS}
% {Author \headeretal: Preparation of Papers for IEEE TRANSACTIONS and JOURNALS}

% \corresp{Corresponding author: First A. Author (e-mail: author@ boulder.nist.gov).}

%%%%%%%%%%%%%%%%%%%%%%%%%%%%%%%%%%%%%%%%%%%%%%%%%%%%
\begin{abstract}
In the incomplete Cholesky-conjugate gradient (ICCG) method, the forward and backward substitutions in the preconditioning step involve sequential dependencies that hinder parallelization on multicore systems.
Algebraic block multi-coloring (ABMC) mitigates this bottleneck through block-wise coloring but requires the number of blocks to be specified in advance.
This study evaluates the Leiden method, a community-detection technique, as an alternative blocking approach that automatically determines blocks reflecting the matrix structure without requiring a predefined block count.
We partition the adjacency graphs of sparse matrices using the Leiden method with modularity and the constant Potts model (CPM) as quality functions and compare the resulting parallel preconditioning with ABMC in terms of iteration count, execution time, and L2 cache efficiency across eight symmetric positive definite matrices.
The results show that Leiden with the CPM was within 10\% of the tuned ABMC in ICCG execution time for five of the eight matrices, including two matrices for which Leiden with CPM was shorter, despite having an average L2 hit rate that was 7.6 percentage points lower.
%%% Revised 20261103
Furthermore, under a cost model that includes ABMC block-count tuning but excludes resolution-parameter selection for Leiden with the CPM, the latter has a lower total cost for all eight matrices when fewer than 25 solves are required.
\end{abstract}


\titlepgskip=-21pt

\maketitle

%%%%%%%%%%%%%%%%%%%%%%%%%%%%%%%%%%%%%%%%%%%%%%%%%%%%
\section{Introduction}
\label{sec:introduction}
Iterative solvers for sparse linear systems are fundamental tools in numerical linear algebra.
In particular, the incomplete Cholesky-conjugate gradient (ICCG) method is widely used in large-scale finite element analysis and fluid simulations.
In ICCG, forward and backward substitutions are used to apply the incomplete Cholesky preconditioner and account for a significant portion of the execution time in each iteration.
Their inherent sequential dependencies, where each element recursively depends on previously computed values, present a major bottleneck for parallelization in multicore environments.

The multi-coloring (MC) method is a conventional approach for parallel preconditioning, in which the adjacency graph derived from the sparse matrix is colored so that independent unknowns belonging to the same color can be processed concurrently.
Although MC exposes parallelism, its fine-grained reordering can reduce data locality and increase cache-related overhead.
To alleviate this bottleneck while preserving data locality, block multi-coloring methods, such as the algebraic block multi-coloring (ABMC) method~\cite{IwashitaAlgebraic2012}, have been developed.
However, the ABMC method requires the number of blocks (or block size) to be specified in advance, and selecting the optimal configuration requires problem-dependent empirical trial and error.

To avoid this problem-dependent block-count tuning, community detection techniques developed for network analysis offer a promising alternative.
By extracting densely connected subgraphs (communities) from the adjacency graph of a sparse matrix and treating them as blocks, densely connected vertices can be grouped within individual blocks without prescribing the number of blocks a priori.
Among quality-function-based community detection algorithms, the Leiden method~\cite{TraagLouvain2019} avoids poorly connected communities and accommodates alternative quality functions, such as modularity~\cite{ClausetFinding2004} and the constant Potts model (CPM)~\cite{TraagNarrow2011}.

However, unlike ABMC, quality-function-based methods do not explicitly enforce uniform block sizes, which may cause parallel load imbalance.
To examine this trade-off, this study applies block multi-coloring to the block structures obtained by both ABMC and Leiden and evaluates their effects on ICCG convergence, parallel execution time, and cache efficiency.
%%% Added 20261103
The comparison is restricted to a common CPU-based shared-memory ICCG framework; performance comparisons with GPU-accelerated implementations are beyond the scope of this study.

The main contributions of this study are summarized as follows:
\begin{itemize}
\item We propose a Leiden-based automated blocking framework for ICCG preconditioning, which avoids the problem-dependent block-count tuning required by ABMC.
\item We evaluate modularity and the CPM as quality functions, analyzing their impacts on block structures, cache efficiency, and solver performance under 32-core execution.
\item Through performance evaluations in a 32-core CPU environment, we demonstrate that the Leiden method with the CPM suppresses the formation of giant blocks and achieves an ICCG execution time difference within 10\% of that of the finely tuned ABMC method for five of the eight test matrices.
\item We evaluate the total cost including the block-count tuning of ABMC and derive break-even numbers of solves, clarifying the trade-off between upfront tuning cost and per-solve ICCG performance.
\end{itemize}

The remainder of this paper is organized as follows.
Section~\ref{sec:related_work} reviews related work on parallel sparse triangular solvers and community detection methods.
Section~\ref{sec:ABMC} describes the ABMC method and the parallel forward and backward substitutions based on block multi-coloring.
Section~\ref{sec:Leiden} introduces the Leiden-based blocking approach and the quality functions used in this study.
Section~\ref{sec:TestMatrices} presents the test matrices and experimental environment.
Sections~\ref{sec:ABMC_Evaluation} and~\ref{sec:Leiden_Evaluation} evaluate the performance of the ABMC- and Leiden-based approaches, respectively.
Section~\ref{sec:Leiden_Evaluation} also compares the total cost of the two approaches, including the block-count tuning of ABMC, and derives the break-even number of solves.
Section~\ref{sec:Scaling} examines their strong-scaling behavior and speedup relative to a sequential baseline, and Section~\ref{sec:Gamma_Evaluation} analyzes the sensitivity of the Leiden-based approaches to the resolution parameters.
Finally, Section~\ref{sec:conclusion} concludes the paper and outlines directions for future work.

%%%%%%%%%%%%%%%%%%%%%%%%%%%%%%%%%%%%%%%%%%%%%%%%%%%%
\section{Related Work}
\label{sec:related_work}
\subsection{Parallel Sparse Triangular Solvers}
Sparse triangular solves are difficult to parallelize because of sequential data dependencies.
Various approaches have therefore been developed to expose parallelism while preserving these dependencies.

Level scheduling is a classical approach for parallel sparse triangular solves~\cite{AndersonSolving1989}.
It models the dependencies as a directed acyclic graph and partitions mutually independent unknowns into levels.
Unknowns within the same level can be computed concurrently, whereas the levels must be processed sequentially.
Subsequent studies have explored multithreaded and many-core sparse triangular solvers using level scheduling, blocking, or hybrid strategies~\cite{BradleyHybrid2016, LiEfficient2020}.

To reduce the synchronization overhead of level-by-level execution, dependency-driven and synchronization-free algorithms have been proposed~\cite{LiuSynchronizationfree2016, LiuFast2017}.
These methods allow each computation to proceed once its predecessors are completed, avoiding global barriers at the cost of runtime dependency management.

Graph-coloring-based methods provide an alternative by reordering unknowns into independent color groups that can be processed in parallel.
Block multi-coloring extends this concept by grouping unknowns into blocks and coloring the block-level adjacency graph.
Compared with fine-grained scheduling approaches, it provides coarser-grained parallelism and improved data locality by processing unknowns within each block sequentially.
In preconditioned iterative methods such as ICCG, the block structure also affects the numerical properties of the preconditioner, requiring a trade-off among parallelism, data locality, synchronization overhead, and convergence.

The ABMC method~\cite{IwashitaAlgebraic2012} applies block multi-coloring to general sparse matrices and is used as the baseline blocking method in this study.
Its blocking, block coloring, and parallel triangular substitutions are described in Section~\ref{sec:ABMC}.

\subsection{Community Detection Methods}
The analysis of large-scale networks has attracted significant attention in computer science~\cite{BroderGraph2000, WattsCollective1998, BarabasiEmergence1999, GirvanCommunity2002}.
Many such networks exhibit community structures, where vertices within the same group are densely connected while connections between groups are sparse.
Community detection methods include divisive~\cite{GirvanCommunity2002}, spectral~\cite{NewmanFinding2006}, dynamic~\cite{PonsComputing2006}, and quality-function-based approaches.
The latter identify communities by optimizing a partition-quality objective without requiring the number of communities in advance.
The Louvain method~\cite{BlondelFast2008} is a representative approach that greedily optimizes modularity~\cite{ClausetFinding2004}, while the Leiden method~\cite{TraagLouvain2019} further refines communities and avoids poorly connected ones.
Both methods can be used to generate blocks from sparse-matrix adjacency graphs without prescribing the number of blocks in advance.
In this study, we focus on Leiden-based blocking because the Leiden method was developed to address limitations of the Louvain method while retaining its ability to automatically determine communities (blocks).

Accordingly, this study compares ABMC and Leiden-based blocking within the same block multi-coloring framework.
The objective is to evaluate whether Leiden-based community detection can provide an effective alternative to the conventional ABMC approach, rather than to comprehensively compare different community detection algorithms.

%%%%%%%%%%%%%%%%%%%%%%%%%%%%%%%%%%%%%%%%%%%%%%%%%%%%
\section{ABMC and Parallel Triangular Solves}
\label{sec:ABMC}
The ABMC method~\cite{IwashitaAlgebraic2012} extends the block multi-coloring method, originally developed for structured grids, to general sparse matrices.
First, an undirected graph is constructed from the nonzero pattern of the coefficient matrix.
Nodes with strong connectivity are subsequently grouped into the same block.
In this blocking process, a seed node is selected from the unassigned nodes, and its neighboring nodes are added to a candidate set until the block reaches a prescribed size.

Next, coloring is performed at the block level, where blocks with mutual dependencies are assigned distinct colors.
Consequently, blocks of the same color become independent, enabling parallel forward and backward substitutions to be executed on a block-by-block basis for each color.

For the forward substitution $Ly=r$, the block colors are processed sequentially to satisfy the data dependencies between colors, whereas the blocks belonging to the same color are processed in parallel.
Synchronization is performed between successive block colors to ensure that all updates for one color are completed before the next color is processed; in the OpenMP implementation, this synchronization is provided by the implicit barrier at the end of each \texttt{omp for} construct.
Within each block, the unknowns are updated sequentially in their prescribed order, which helps preserve data locality.
This combination enhances both cache efficiency and the quality of the preconditioner.
Here, the quality of the preconditioner refers to how closely the preconditioning matrix constructed via the incomplete Cholesky factorization with zero fill-in, or IC(0), approximates the original coefficient matrix.
Algorithm~\ref{alg:parallel_forward} summarizes this parallel forward substitution, where $r_i$ and $y_i$ denote the $i$th components of the right-hand-side and intermediate vectors, respectively, and $L_{ij}$ denotes the $(i,j)$ entry of the incomplete Cholesky factor $L$.
For each block color, the blocks are distributed among OpenMP threads.
The backward substitution is performed analogously by processing the block colors and the unknowns within each block in reverse order.

\begin{algorithm}[ht]
\caption{Parallel forward substitution with block multi-coloring}
\label{alg:parallel_forward}
\begin{algorithmic}[1]
\For{$c = 1, \ldots, n_c$}
    \State \textbf{parallel for} each block $B$ with color $c$
    \For{each row $i$ in $B$ in ascending order}
        \State $s \gets r_i$
        \For{each $j < i$ such that $L_{ij} \neq 0$}
            \State $s \gets s - L_{ij} y_j$
        \EndFor
        \State $y_i \gets s / L_{ii}$
    \EndFor
    \State \textbf{end parallel for}
    \State Synchronize threads
\EndFor
\end{algorithmic}
\end{algorithm}

The ABMC method requires the block size as an input parameter, which significantly influences the performance of the ICCG method.
A larger block size typically improves cache efficiency and convergence behavior, whereas a smaller block size increases the degree of parallelism.
Because the optimal configuration depends heavily on the specific problem, determining it requires problem-dependent tuning.

%%%%%%%%%%%%%%%%%%%%%%%%%%%%%%%%%%%%%%%%%%%%%%%%%%%%
\section{Leiden-Based Blocking and Quality Functions}
\label{sec:Leiden}
The Leiden method~\cite{TraagLouvain2019} improves upon the Louvain method by introducing a refinement phase to avoid poorly connected communities.

Specifically, after the quality function is optimized through the local movement of vertices, the Leiden method applies a refinement phase to the resulting partition.
During this phase, communities are sub-partitioned to guarantee high internal connectivity before the aggregated (contracted) graph is constructed.
This multi-stage procedure is then repeated iteratively.
Consequently, the Leiden method inherits the fast multilevel optimization framework of the Louvain method while ensuring connectivity within the extracted communities.

Various quality functions, such as modularity and the CPM~\cite{TraagNarrow2011}, can be employed within the Leiden method.
Modularity~\cite{ClausetFinding2004} quantifies the difference between the actual number of edges within communities and the expected number of such edges in a random null model that preserves the degree distribution.
It is formally defined as
\begin{equation}
  Q_{\mathrm{MD}} = \frac{1}{2m} \sum_{i,j} \left( A_{ij} - \gamma_{\mathrm{MD}} \frac{k_i k_j}{2m} \right) \delta(c_i,c_j),
  \label{eq:modularity}
\end{equation}
where $A_{ij}$ is the $(i,j)$th element of the adjacency matrix, $k_i$ is the degree of vertex $i$, $m$ is the total number of edges, $\gamma_{\mathrm{MD}}$ is the resolution parameter, and $c_i$
denotes the community to which vertex $i$ belongs.
The function $\delta(c_i,c_j)$ equals 1 if vertices $i$ and $j$ belong to the same community and 0 otherwise.
Maximizing this metric yields a partition where edges are densely concentrated within communities and relatively sparse between them.

In contrast, the CPM does not rely on a null model derived from a random graph.
Its quality function is formulated as
\begin{equation}
Q_{\mathrm{CP}} = \frac{1}{2m} \sum_{i,j} \left( A_{ij} - \gamma_{\mathrm{CP}} \right) \delta(c_i,c_j),
\label{eq:cpm}
\end{equation}
where $\gamma_{\mathrm{CP}}$ is the resolution parameter.
Unlike modularity, which uses the null-model-dependent term $\gamma_{\mathrm{MD}} k_i k_j/(2m)$, the CPM uses the constant $\gamma_{\mathrm{CP}}$ as a threshold for within-community edge density.
Thus, $\gamma_{\mathrm{CP}}$ directly controls the resolution of the resulting partition, with larger values generally favoring smaller communities.
Hereafter, \texttt{LeidenMD} and \texttt{LeidenCP} denote the Leiden method based on modularity maximization and CPM maximization, respectively.

In this study, we apply the block (community) partitions obtained via these two distinct quality functions to parallel preconditioning for the ICCG method.
We then comprehensively evaluate the impacts of these quality functions on the resulting number of blocks, block-size distribution, number of block colors, number of ICCG iterations, and parallel execution time.

%%%%%%%%%%%%%%%%%%%%%%%%%%%%%%%%%%%%%%%%%%%%%%%%%%%%
\section{Test Matrices and Experimental Environment}
\label{sec:TestMatrices}
This section describes the test matrices and the computational setup used in our evaluation.

\begin{table*}[htb]
\centering
\caption{Characteristics of sparse test matrices}
\label{tab::FloridaMat}
\begin{tabular}{l|r|r|r|l|l}
\hline
\multicolumn{1}{c|}{Matrix} & Size ($n$) & Nonzeros ($n_z$) & ANZR & \multicolumn{1}{c|}{SCN} & \multicolumn{1}{c}{Problem type}
\rule{0pt}{1.0em} 
\\
\hline
\texttt{parabolic\_fem} & 525,825   & 3,674,625  & 6.99  & $2.10 \times 10^5$ & CFD \\
\texttt{apache2}        & 715,176   & 4,816,870  & 6.74  & $3.07 \times 10^6$ & Structural \\
\texttt{Emilia\_923}    & 923,136   & 40,373,538 & 43.7 & \multicolumn{1}{c|}{---} & Structural \\
\texttt{ecology2}       & 999,999   & 4,995,991  & 5.00  & $6.53 \times 10^7$ & 2D/3D \\
\texttt{thermal2}       & 1,228,045 & 8,580,313  & 6.99  & $4.85 \times 10^6$ & Thermal \\
\texttt{Geo\_1438}      & 1,437,960 & 60,236,322 & 41.9 & $1.14 \times 10^{13}$ & Structural \\
\texttt{Hook\_1498}     & 1,498,023 & 59,374,451 & 39.6 & $3.60 \times 10^6$ & Structural \\
\texttt{G3\_circuit}    & 1,585,478 & 7,660,826  & 4.83  & $1.47 \times 10^7$ & Circuit simulation \\
\hline
\end{tabular}
\vspace{1mm}

\begin{minipage}{0.98\textwidth}
\footnotesize
\centering
For \texttt{Emilia\_923}, the SCN could not be reliably estimated because the smallest Ritz value did not converge.
\end{minipage}
\end{table*}

We selected eight real, symmetric positive definite matrices from the SuiteSparse Matrix Collection~\cite{DavisUniversity2011}.
The matrices were selected to cover a range of problem sizes and sparsity characteristics.
For all test matrices, the IC(0) factorization was successfully completed without breakdown after applying each of the block multi-coloring schemes considered in this study.
As summarized in Table~\ref{tab::FloridaMat}, these matrices originate from diverse application domains, including computational fluid dynamics (CFD), structural analysis, thermal analysis, circuit simulation, and grid-based problems, and cover a wide range of average nonzeros per row ($\mathrm{ANZR} = n_z / n$).
The spectral condition number ($\mathrm{SCN}$) was included as an indicator of the numerical characteristics of each test matrix and was estimated from the CG iteration history using the Lanczos-based approach described in \cite{SaadIterative2003}.

The hardware and software specifications of the computational environment utilized in our experiments are as follows.
The system featured an AMD EPYC 7543 processor with 32 cores, 64 KB of L1 cache per core, 512 KB of L2 cache per core, and a shared 256 MB of L3 cache in total (32MB per CCX).
The main memory consisted of 128 GB of DDR4 RAM, configured as eight 16-GB modules.
The operating system was Ubuntu 22.04 LTS.

The ICCG method was implemented in C++, with Intel oneMKL 2025.2 utilized as the BLAS library.
The source code was compiled using g++ version 11.4.0 with the \mbox{-O3} optimization option.
OpenMP \texttt{parallel for} directives with static scheduling were used for parallelizable loops.
No explicit thread affinity policy was specified for OpenMP in the main experiments.
As a sensitivity analysis, thread affinity was evaluated for all eight matrices by setting \texttt{OMP\_PLACES=cores} with \texttt{OMP\_PROC\_BIND=close} or \texttt{spread}.
Compared with the default settings, explicit thread binding changed the execution time by up to approximately 7\% for ABMC, 6\% for \texttt{LeidenMD}, and 2\% for \texttt{LeidenCP}, with the direction of the change depending on the matrix and method.
Neither \texttt{OMP\_PROC\_BIND=close} nor \texttt{spread} consistently improved performance.
%%% Added 20261103
This sensitivity analysis quantifies thread-binding effects but does not identify their underlying hardware or operating-system mechanisms.
The relationship between thread placement and block structure requires further investigation.

For the linear system $Ax=b$, the right-hand-side vector $b$ was set to a vector of all ones for all test matrices to ensure consistent experimental conditions.
The convergence criterion was defined by the relative residual norm satisfying $\lVert r \rVert_2 / \lVert b \rVert_2 \le 1.0 \times 10^{-8}$, where $r$ denotes the residual vector.
Both the execution time and the number of iterations were evaluated as averages over 10 independent runs.

The test matrices used in this study are publicly available from the SuiteSparse Matrix Collection~\cite{DavisUniversity2011}.
The source code used in this study and the experimental data generated in this work are available from the corresponding author upon reasonable request.

%%%%%%%%%%%%%%%%%%%%%%%%%%%%%%%%%%%%%%%%%%%%%%%%%%%%
\section{Performance Evaluation Using the ABMC Method}
\label{sec:ABMC_Evaluation}
\subsection{Execution Performance and Optimal Number of Blocks}
\begin{figure}[htb]
\centering
\includegraphics[width=\linewidth]{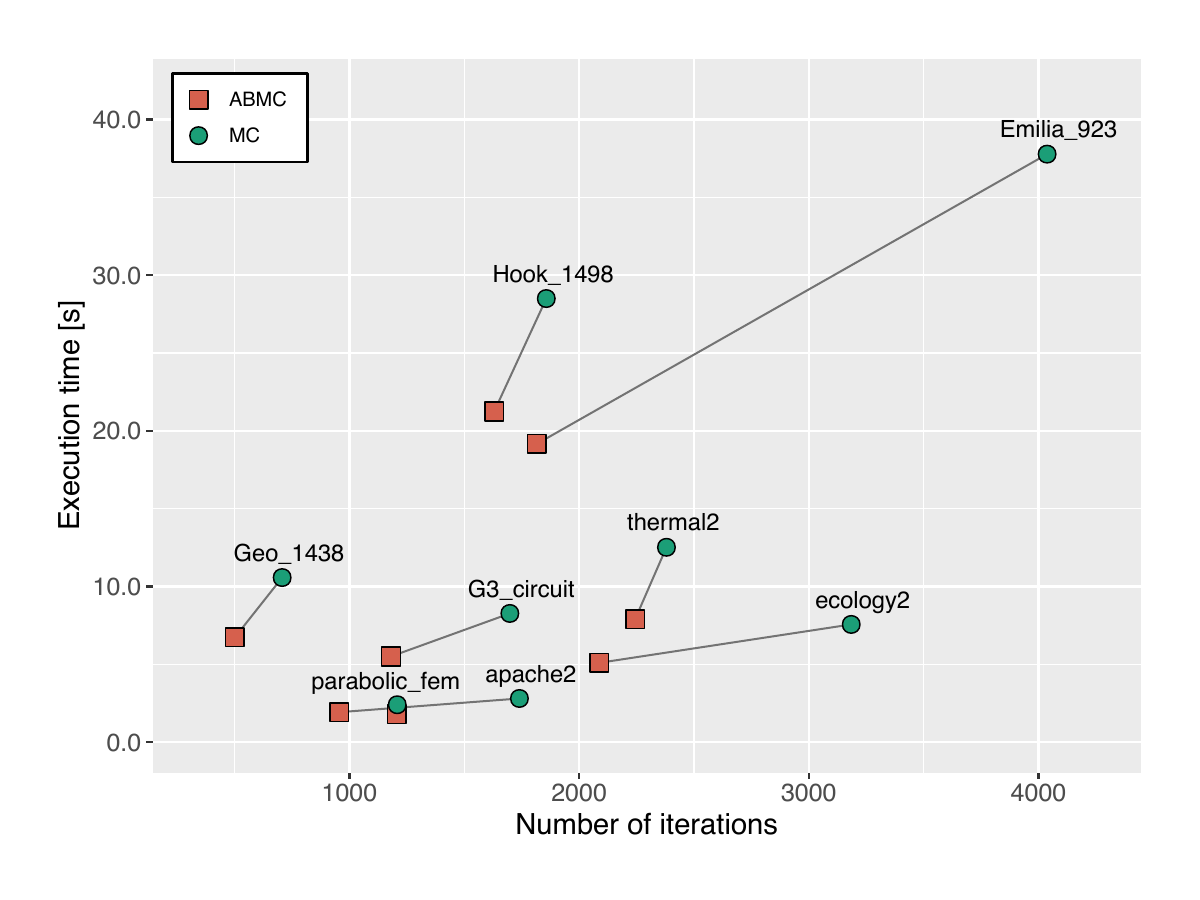}
\caption{Scatter plot of average execution time versus average number of iterations for the best-performing ABMC configuration and conventional MC}
\label{fig::Fig1}
\end{figure}

Fig.~\ref{fig::Fig1} shows the execution results of the ICCG method with the ABMC and conventional MC methods used for parallel preconditioning.
In the evaluation of the ABMC method, block partitioning was performed according to the procedure described in Section~\ref{sec:ABMC}, followed by block-level coloring using the greedy coloring algorithm~\cite{SaadIterative2003}.
For the conventional MC method, the same greedy coloring algorithm was applied directly to the adjacency graph at the node level without block partitioning.
The figure presents scatter plots illustrating the relationship between the average execution time and the average number of iterations of the ICCG method across the test matrices, with the ABMC and MC results for each matrix connected by a line.

\begin{table*}[htb]
\begin{center}
\caption{Best-performing ABMC configurations and comparison with conventional MC for the tested matrices}
\label{tab::BestNB_MC}
\vspace{0.2em}
\begin{tabular}{l|r|r|c|c||r|c|c}
\hline
\multicolumn{1}{c|}{\multirow{2}{*}{Matrix} } & \multicolumn{4}{c||}{ABMC} & \multicolumn{3}{c}{MC} \\ \cline{2-8}
\rule[-0.8ex]{0pt}{3.0ex}
& \multicolumn{1}{c|}{$n_b$} & \multicolumn{1}{c|}{$n_c$} & \multicolumn{1}{c|}{Execution time [s]} & \multicolumn{1}{c||}{No. iterations} &
\multicolumn{1}{c|}{$n_c$} & \multicolumn{1}{c|}{Execution time [s]} & \multicolumn{1}{c}{No. iterations}\\
\hline
\texttt{parabolic\_fem} &   128 &  6 & 1.79 $\pm$ 0.019 & 1,206.5 &  5 & 2.40 $\pm$ 0.067 & 1,208.0 \\
\texttt{apache2}        &   128 &  5 & 1.92 $\pm$ 0.022 &   955.0 &  3 & 2.80 $\pm$ 0.086 & 1,740.0 \\
\texttt{Emilia\_923}    &    64 &  8 & 19.2 $\pm$ 0.31  & 1,815.9 & 36 & 37.8 $\pm$ 0.18 & 4,037.3 \\
\texttt{ecology2}       &   128 &  5 & 5.09 $\pm$ 0.072 & 2,087.0 &  2 & 7.56 $\pm$ 0.048 & 3,184.6 \\
\texttt{thermal2}       & 1,024 &  6 & 7.89 $\pm$ 0.062 & 2,243.6 &  6 & 12.5 $\pm$ 0.097 & 2,380.1 \\
\texttt{Geo\_1438}      &   128 &  8 & 6.73 $\pm$ 0.060 &   501.2 & 36 & 10.6 $\pm$ 0.041 & 707.0 \\
\texttt{Hook\_1498}     &   512 & 11 & 21.2 $\pm$ 0.057 & 1,630.1 & 30 & 28.5 $\pm$ 0.19 & 1,857.0 \\
\texttt{G3\_circuit}    &   512 & 10 & 5.50 $\pm$ 0.081 & 1,181.0 &  4 & 8.26 $\pm$ 0.045 & 1,698.3 \\
\hline
\end{tabular}
\end{center}
\end{table*}

The ABMC method employs the number of blocks $n_b$, or equivalently the block size $s = \lceil n/n_b \rceil$, as a critical runtime parameter that significantly influences the performance of the ICCG method.
To investigate this dependency, we evaluated the execution time of the ICCG method for $n_b$ values of 32, 64, 128, 256, 512, and 1024.
Table~\ref{tab::BestNB_MC} summarizes the optimal value of $n_b$ that achieved the shortest ICCG execution time for each test matrix, together with the corresponding number of block colors $n_c$, execution time reported as the mean $\pm$ standard deviation over 10 runs, and the average number of iterations over the same 10 runs.
% 20261103
For comparison, the same quantities are reported for the conventional MC method under the same experimental conditions.
The non-integer iteration counts result from averaging over 10 runs, as the iteration count may vary slightly between runs because of floating-point rounding differences caused by the non-deterministic reduction order in OpenMP parallelization.
The execution results of the optimized ABMC configurations and the conventional MC method are plotted in Fig.~\ref{fig::Fig1}.

ABMC generally required fewer ICCG iterations than conventional MC, while the difference was negligible for \texttt{parabolic\_fem}.
ABMC also achieved shorter execution times than conventional MC for all eight test matrices.
The reduction in execution time is modest for \texttt{parabolic\_fem} and \texttt{Hook\_1498}, but more pronounced for the other six matrices.
This observation is consistent with the results reported by Iwashita et al.~\cite{IwashitaAlgebraic2012} and supports the use of ABMC as the representative conventional baseline in the subsequent comparisons.

\subsection{L2 Cache Efficiency}
To evaluate L2 cache efficiency during the execution of the ICCG method, we collected the following hardware performance events utilizing the Linux \texttt{perf} tool:
\begin{itemize}
\item \texttt{l2\_cache\_accesses\_from\_dc\_misses}  \\ ($\mathrm{L2}_{\mathrm{access}}$)
\item \texttt{l2\_cache\_hits\_from\_dc\_misses}  ($\mathrm{L2}_{\mathrm{hit}}$)
\item \texttt{l2\_cache\_misses\_from\_dc\_misses}  ($\mathrm{L2}_{\mathrm{miss}}$)
\end{itemize}
Based on these metrics, the L2 hit rate and L2 miss rate were respectively defined as
%  $\text{L2 hit rate} = \mathrm{L2}_{\mathrm{hit}} / \mathrm{L2}_{\mathrm{access}}$ and $\text{L2 miss rate} = \mathrm{L2}_{\mathrm{miss}} / \mathrm{L2}_{\mathrm{access}}$, respectively.
\begin{equation}
\begin{aligned}
R_{\mathrm{hit}}
& =
\frac{\mathrm{L2}_{\mathrm{hit}}}
     {\mathrm{L2}_{\mathrm{access}}}, \\
R_{\mathrm{miss}}
& =
\frac{\mathrm{L2}_{\mathrm{miss}}}
     {\mathrm{L2}_{\mathrm{access}}}.
\end{aligned}
\label{eq:L2_cache_efficiency}
\end{equation}
Following the evaluation methodology established in~\cite{IwashitaAlgebraic2012}, this study focuses exclusively on L2 cache efficiency. 
Because the L3 cache on the target AMD EPYC processor is shared among multiple cores within a core complex (CCX), L3 metrics are heavily affected by inter-thread contention.
In contrast, L2 metrics provide a more direct and reliable indicator of the core-level data locality achieved by the preconditioning algorithms.
%%% Added 20261103
These metrics characterize L2 cache locality rather than DRAM bandwidth utilization or memory access latency, and do not establish whether execution is memory-bandwidth-bound.
Table~\ref{tab::ABMC_ICCG_cache} summarizes the measured L2 cache efficiency profiles for the ABMC method.

\begin{table}[htb]
\begin{center}
\caption{L2 cache efficiency of ABMC-based ICCG}
\label{tab::ABMC_ICCG_cache}
\begin{tabular}{l|c|c}
\hline
\multicolumn{1}{c|}{Matrix} & $R_{\mathrm{hit}}$ & $R_{\mathrm{miss}}$ \\
\hline
\texttt{parabolic\_fem}  & 0.87 & 0.065 \\
\texttt{apache2}         & 0.85 & 0.060 \\
\texttt{Emilia\_923}     & 0.79 & 0.040 \\
\texttt{ecology2}        & 0.86 & 0.052 \\
\texttt{thermal2}        & 0.88 & 0.064 \\
\texttt{Geo\_1438}       & 0.74 & 0.049 \\
\texttt{Hook\_1498}      & 0.69 & 0.10  \\
\texttt{G3\_circuit}     & 0.83 & 0.082 \\
\hline
\end{tabular}
\end{center}
\end{table}

%%%%%%%%%%%%%%%%%%%%%%%%%%%%%%%%%%%%%%%%%%%%%%%%%%%%
\subsection{Impact of ABMC Blocks on Parallel Preconditioning Performance}
We now discuss the performance of the ICCG method with parallel preconditioning using the ABMC method across the eight test matrices.
First, as demonstrated in Table~\ref{tab::BestNB_MC}, the optimal number of blocks $n_b$ varies significantly depending on the matrix.
This variation indicates that the appropriate number of blocks in the ABMC method is determined not only by the matrix size or ANZR but also by the underlying nonzero structure and local connectivity patterns, strongly supporting the observation in~\cite{IwashitaAlgebraic2012} that the optimal configuration is highly problem-dependent.

The performance profiles plotted in Fig.~\ref{fig::Fig1} and summarized in Table \ref{tab::BestNB_MC} reveal that the execution time is dictated by a complex interplay between the number of iterations and cache efficiency.
For instance, low to moderate ANZR matrices such as \texttt{parabolic\_fem}, \texttt{apache2}, \texttt{ecology2}, \texttt{thermal2}, and \texttt{G3\_circuit} achieve high L2 hit rates of 0.83 or higher, demonstrating that block partitioning successfully enhances memory access locality.
%%% Modified 20261105
Conversely, matrices from structured analysis with high ANZR values, such as \texttt{Emilia\_923}, \texttt{Geo\_1438}, and \texttt{Hook\_1498}, incur massive computational workloads per iteration.
% Conversely, structured matrices with high ANZR values, such as \texttt{Emilia\_923}, \texttt{Geo\_1438}, and \texttt{Hook\_1498}, incur massive computational workloads per iteration.
While \texttt{Geo\_1438} mitigates this workload through a remarkably small number of iterations ($\approx 500$), \texttt{Emilia\_923} and \texttt{Hook\_1498} exhibit increased execution times.
The relatively low L2 hit rate for \texttt{Hook\_1498} suggests that limited cache reuse may contribute to its longer execution time.

Among the low-to-moderate ANZR matrices above, \texttt{apache2} and \texttt{ecology2} both require only five block colors and achieve similar L2 hit rates of 0.85 and 0.86, respectively.
However, \texttt{ecology2} requires a substantially larger number of ICCG iterations than \texttt{apache2}, which leads to a longer execution time despite its high L2 hit rate.
This comparison indicates that favorable cache behavior alone does not guarantee short execution time when the convergence behavior is poor.

Overall, the optimal $n_b$ is strongly problem-dependent, motivating an automated blocking framework.

%%%%%%%%%%%%%%%%%%%%%%%%%%%%%%%%%%%%%%%%%%%%%%%%%%%%
\section{Performance Evaluation Using The Leiden Method}
\label{sec:Leiden_Evaluation}
\subsection{Leiden Configuration and Seed Sensitivity}
Next, we evaluate the performance of the ICCG method with the Leiden method used for block partitioning in parallel preconditioning.
For the Leiden method, we utilized the \texttt{igraph\_community\_leiden\_simple} function implemented in \texttt{igraph} version 1.0.1~\cite{CsardiIgraph2006}.
The two quality functions to be maximized are modularity and the CPM.
The resolution parameter for modularity, $\gamma_{\mathrm{MD}}$, was set to its default value of 1.0.
The resolution parameter for the CPM, $\gamma_{\mathrm{CP}}$, was configured as $1.0 \times 10^{-3}$, which demonstrated high performance in preliminary experiments.
The number of Leiden iterations was set to two, and the refinement parameter \texttt{beta} was set to 0.01.
The sensitivity of the Leiden-based blocking to $\gamma_{\mathrm{MD}}$ and $\gamma_{\mathrm{CP}}$ is examined in Section~\ref{sec:Gamma_Evaluation}.

The generated blocks were colored utilizing the greedy coloring algorithm and then applied to the parallel preconditioning to evaluate the execution performance of the ICCG method.
%%% Added 20261103
Using the same greedy coloring algorithm provides a common basis for comparing the blocking methods; alternative coloring strategies are left for future work.
%%%
Unlike the ABMC method, the Leiden method inherently generates blocks of non-uniform sizes, meaning that the block ordering can also affect the overall performance.
To optimize data reference locality, this study sorts the blocks in descending order of size, assigning smaller indices to larger blocks.
This reordering places the unknowns of larger blocks with strong internal connections into contiguous index ranges, thereby concentrating the nonzero elements near the main diagonal and improving cache reuse during forward and backward substitutions.

Since the Leiden method involves stochastic operations, we also examined the sensitivity of the generated blocks and ICCG performance to the random seed.
%%% modified { -> \{ in 20261105
To assess the stochastic variability of the Leiden method, five random seeds, $\{9,31,42,58,74\}$, were tested.
Table~\ref{tab::Leiden_Seed_Variation} summarizes the variations in the generated block structures and ICCG performance.
Here, $\Delta n_b$, $\Delta N_{\mathrm{iter}}$, and $\Delta T$ denote the relative ranges of the number of blocks, the ICCG iteration count, and the execution time, respectively, defined as $(x_{\max}-x_{\min})/\bar{x}\times100$ (\%).
The quantities $n_c^{\min}$ and $n_c^{\max}$ denote the minimum and maximum numbers of block colors observed across the five seeds, respectively.
For each seed, the execution time was averaged over 10 runs, and $\Delta T$ was calculated from these five mean execution times.

\begin{table*}[htb]
\begin{center}
\small
\caption{Variations in Leiden-based block structures and ICCG performance across five random seeds}
\label{tab::Leiden_Seed_Variation}
\vspace{0.3em}
\begin{tabular}{l|r|c|c|c||r|c|c|c}
\hline
\multicolumn{1}{c|}{ \multirow{2}{*}{Matrix} } & \multicolumn{4}{c||}{\texttt{LeidenMD}} & \multicolumn{4}{c}{\texttt{LeidenCP}} \\ \cline{2-9}
\rule[-0.8ex]{0pt}{3.0ex}
& \multicolumn{1}{c|}{$\Delta n_b$} & $n_c^{\text{min}} / n_c^{\text{max}}$ & \multicolumn{1}{c|}{$\Delta N_{\text{iter}}$} & \multicolumn{1}{c||}{$\Delta T$} & \multicolumn{1}{c|}{$\Delta n_b$} & $n_c^{\text{min}} / n_c^{\text{max}}$ & \multicolumn{1}{c|}{$\Delta N_{\text{iter}}$} & \multicolumn{1}{c}{$\Delta T$} \\ \hline
\texttt{parabolic\_fem}
&  11.3 & 5 / 6 &  1.25 & 12.3 & 0.920 & 6 /  6 &  6.05 & 6.86 \\
\texttt{apache2}
&  7.12 & 5 / 6 &  2.09 & 4.50 &  1.76 & 8 /  8 &  1.56 & 3.32 \\
\texttt{Emilia\_923}
&  15.5 & 7 / 8 &  43.6 & 37.3 &  3.16 & 9 /  9 &  41.1 & 40.1 \\
\texttt{ecology2}
&  4.93 & 5 / 6 &  1.16 & 12.3 &  1.26 & 6 /  6 &  2.49 & 2.82 \\
\texttt{thermal2}
& 0.455 & 5 / 5 &  1.34 & 16.6 & 0.374 & 6 /  6 &  3.40 & 3.44 \\
\texttt{Geo\_1438}
&  8.02 & 5 / 6 &  1.22 & 6.75 &  2.53 & 8 /  9 & 0.584 & 2.26 \\
\texttt{Hook\_1498}
&  11.2 & 6 / 6 & 0.593 & 7.53 &  1.29 & 9 / 10 & 0.553 & 2.05 \\
\texttt{G3\_circuit}
& 0.867 & 5 / 5 &  2.58 & 12.2 & 0.276 & 7 /  8 & 0.745 & 1.14 \\
\hline
\end{tabular}
\end{center}
\end{table*}

Table~\ref{tab::Leiden_Seed_Variation} shows that the stochastic variability was generally limited, particularly for \texttt{LeidenCP}.
Except for \texttt{Emilia\_923}, the relative variation in the ICCG execution time for \texttt{LeidenCP} was at most 6.9\%, while the variation in the iteration count was at most 6.1\%.
In contrast, \texttt{Emilia\_923} exhibited much larger variations, with the iteration count varying by approximately 41--44\% and the execution time by approximately 37--40\% for both Leiden-based methods.
This result indicates that the ICCG performance for \texttt{Emilia\_923} is particularly sensitive to the stochastic variation of the generated blocks.
Based on these results, a fixed random seed of 42 was used in the following experiments; this value was chosen for reproducibility, not to optimize performance.

\subsection{Execution Performance}
\begin{figure}[htb]
\centering
\includegraphics[width=\linewidth]{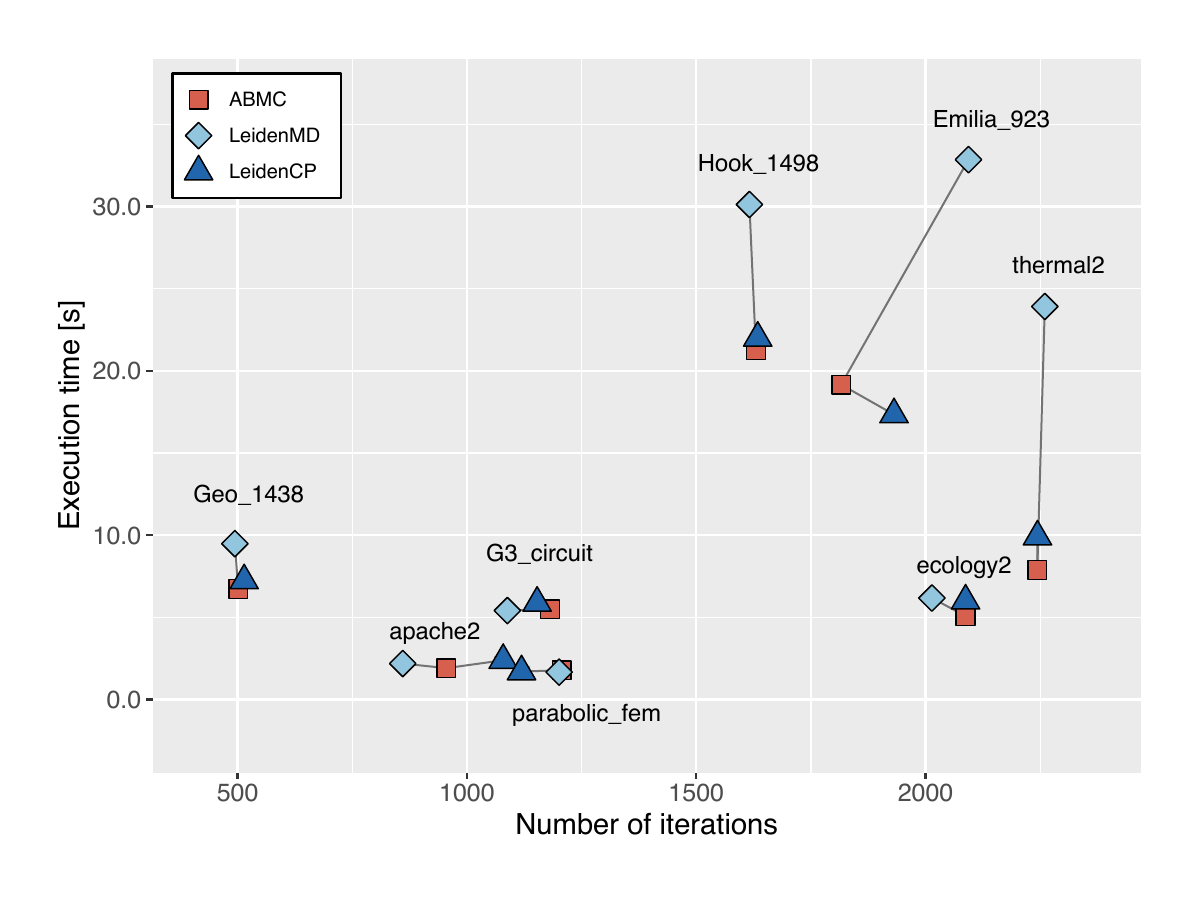}
\caption{Scatter plot of average execution time versus average number of iterations for Leiden- and ABMC-based methods}
\label{fig::Scatters}
\end{figure}

Fig.~\ref{fig::Scatters} presents the execution performance of the ICCG method with the Leiden method used for parallel preconditioning.
For comparison, the optimized results of the ABMC method from Fig.~\ref{fig::Fig1} are also plotted, with data points corresponding to the same matrix connected by lines.
Table~\ref{tab::Leiden_NB_NC} summarizes the number of blocks generated by the Leiden method for each matrix, along with the corresponding number of block colors and execution times.
The execution times are reported as the mean $\pm$ standard deviation over 10 runs, and the iteration counts are reported as the corresponding averages.

%%%%%%%%%%%%%%%%%%%%%%%%%%%%%%%%%%%%%%%%%%%%%%%%%%%%

\begin{table*}[htb]
\begin{center}
\small
\caption{$n_b$ values determined by Leiden-based blocking, corresponding $n_c$ values, and execution times for tested matrices}
\label{tab::Leiden_NB_NC}
\vspace{0.3em}
\begin{tabular}{l|r|c|c|c||r|c|c|c}
\hline
\multicolumn{1}{c|}{ \multirow{2}{*}{Matrix} } & \multicolumn{4}{c||}{\texttt{LeidenMD}} & \multicolumn{4}{c}{\texttt{LeidenCP}} \\ \cline{2-9}
& \multicolumn{1}{c|}{$n_b$} & $n_c$ & \multicolumn{1}{c|}{Execution time [s]} & \multicolumn{1}{c||}{No. iterations} & \multicolumn{1}{c|}{$n_b$} & $n_c$ & \multicolumn{1}{c|}{Execution time [s]} &  \multicolumn{1}{c}{No. iterations} \rule{0pt}{1.0em}
\\ \hline
\texttt{parabolic\_fem}
& 101   & 5 & 1.68 $\pm$ 0.0046 & 1,201.0 & 2,277 &  6 & 1.71 $\pm$ 0.0018 & 1,119.0 \\
\texttt{apache2}
&  69   & 5 & 2.20 $\pm$ 0.0083 &   860.0 & 2,253 &  8 & 2.40 $\pm$ 0.016  & 1,079.0 \\
\texttt{Emilia\_923}
&  39   & 8 & 32.9 $\pm$    1.0 & 2,093.6 &   448 &  9 & 17.3 $\pm$ 0.046  & 1,931.2 \\
\texttt{ecology2}
& 144   & 6 & 6.18 $\pm$  0.047 & 2,013.9 & 6,366 &  6 & 6.02 $\pm$ 0.067  & 2,087.2 \\
\texttt{thermal2}
& 1,099 & 5 & 23.9 $\pm$ 0.11   & 2,260.0 & 6,404 &  6 & 9.91 $\pm$ 0.037  & 2,244.0 \\
\texttt{Geo\_1438}
& 37    & 6 & 9.49 $\pm$ 0.21   &   494.0 &   721 &  8 & 7.25 $\pm$ 0.053  &   514.2 \\
\texttt{Hook\_1498}
& 46    & 6 & 30.1 $\pm$ 0.60   & 1,615.9 &   779 & 10 & 22.0 $\pm$ 0.057  & 1,634.0 \\
\texttt{G3\_circuit}
& 116   & 5 & 5.42 $\pm$ 0.017  & 1,088.2 & 9,777 &  7 & 5.88 $\pm$ 0.027  & 1,153.0 \\
\hline
\end{tabular}
\end{center}
\end{table*}
%%%%%%%%%%%%%%%%%%%%%%%%%%%%%%%%%%%%%%%%%%%%%%%%%%%

Fig.~\ref{fig::Scatters} and Tables~\ref{tab::BestNB_MC} and \ref{tab::Leiden_NB_NC} show clear performance differences between \texttt{LeidenMD} and \texttt{LeidenCP}.
For \texttt{LeidenMD}, shorter execution times than those of the ABMC method were obtained for \texttt{parabolic\_fem} and \texttt{G3\_circuit}, whereas longer execution times were observed for the other six matrices.
\texttt{parabolic\_fem} and \texttt{G3\_circuit} are characterized by relatively low to moderate ANZR values and grid-like or sparse graph structures, for which the coarse communities generated by modularity maximization do not necessarily cause severe sequential bottlenecks.
For \texttt{thermal2}, the execution time increased significantly even though the average number of iterations remained almost identical to that of the ABMC method.
This discrepancy suggests that under the \texttt{LeidenMD} partitioning, the severe performance degradation arises not from the convergence behavior itself, but rather from block-size imbalance, an increase in sequential processing overhead within massive blocks, or parallel load imbalance.

\subsection{Block-Size and Load-Imbalance Analysis}
As shown in Table~\ref{tab::Leiden_NB_NC}, \texttt{LeidenCP} achieved shorter execution times than \texttt{LeidenMD} for five of the eight matrices.
The improvement was particularly pronounced for \texttt{Emilia\_923}, \texttt{thermal2}, \texttt{Geo\_1438}, and \texttt{Hook\_1498}.
As detailed in Table~\ref{tab::Leiden_NB_NC}, this performance gain is driven by the fact that the CPM function produces finer block partitions than those obtained with modularity maximization.
This finer partitioning reduces the sequential workload within oversized blocks.

However, for \texttt{G3\_circuit}, \texttt{LeidenMD} exhibits a shorter execution time than that of \texttt{LeidenCP}, demonstrating that finer partitioning via the CPM is not universally advantageous.
Thus, the choice of quality function involves trade-offs among block granularity, the number of block colors, load balance, and data locality.

Tables~\ref{tab::BestNB_MC} and \ref{tab::Leiden_NB_NC} show that \texttt{LeidenMD} generates fewer blocks than those for the optimal configurations of the ABMC method, except for \texttt{ecology2} and \texttt{thermal2}.
For example, \texttt{Emilia\_923}, \texttt{Geo\_1438}, and \texttt{Hook\_1498} are partitioned into only 39, 37, and 46 blocks, respectively, which are significantly smaller than the best $n_b$ values for the ABMC method.
Additionally, the number of block colors remains restricted to approximately 5 to 8.
From the standpoint of synchronization costs alone, this small number of block colors should be advantageous.
However, as detailed later in Table~\ref{tab::load_imbalance_stats}, a smaller number of blocks naturally leads to larger individual blocks, thereby increasing the sequential processing within each block.
Consequently, a reduced number of block colors does not automatically guarantee a shorter execution time.

In contrast, \texttt{LeidenCP} generates significantly more blocks than those for \texttt{LeidenMD} across the test matrices, resulting in much finer partitions.
Nevertheless, the number of block colors remains relatively small, approximately 6 to 10 in most cases.
This result indicates that CPM-based partitioning increases the number of parallel execution units without causing a proportional increase in the synchronization overhead associated with block coloring.

A key advantage of deploying the Leiden method as a blocking mechanism for parallel preconditioning is that it avoids explicitly prescribing the number of blocks or block sizes.
However, the method is not entirely parameter-free, because the resolution parameter of the quality function, such as $\gamma_{\mathrm{MD}}$ or $\gamma_{\mathrm{CP}}$, still influences the resulting block structure.
Moreover, the resulting block sizes are inherently non-uniform, which can lead to load imbalance during parallel execution.
%%%%%%%%%%%%%%%%%%%%%%%%%%%%%%%%%%%%%%%%%%%%%%%%%%%%%%%%%%%%%%

\begin{table*}[htb]
\caption{Load imbalance statistics and 32-thread speedup for representative blockings.
$B_{\max}/B_{\mathrm{avg}}$ denotes the ratio of the maximum block size to the average block size,
$C_{\max}/C_{\mathrm{avg}}$ denotes the ratio of the maximum color-wise node count to the average color-wise node count,
and $S_{32}$ denotes the speedup at 32 threads relative to single-thread execution of the same method.}
\label{tab::load_imbalance_stats}
\centering
\small
\begin{tabular}{l|l|r|r|c|r|c|c|c}
\hline
\multicolumn{1}{c|}{Matrix} & \multicolumn{1}{c|}{Method}
& \multicolumn{1}{c|}{$n_b$} & \multicolumn{1}{c|}{$n_c$}
& \multicolumn{1}{c|}{$G$}
& $B_{\max}$
& $B_{\max}/B_{\mathrm{avg}}$
& $C_{\max}/C_{\mathrm{avg}}$
& $S_{32}$ \\
\hline
\texttt{parabolic\_fem}
 & ABMC              &   128 & 6 & 0.00 &  4,109 & 1.00 & 1.45 &  8.48 \\
 & \texttt{LeidenMD} &   101 & 5 & 0.04 &  6,295 & 1.21 & 1.41 &  8.95 \\
 & \texttt{LeidenCP} & 2,277 & 6 & 0.05 &    407 & 1.76 & 1.41 & 10.4 \\
\hline
\texttt{apache2}
 & ABMC              &   128 & 5 & 0.00 &  5,588 & 1.00 & 1.17 & 11.6 \\
 & \texttt{LeidenMD} &    69 & 5 & 0.14 & 16,639 & 1.61 & 1.35 &  8.36 \\
 & \texttt{LeidenCP} & 2,253 & 8 & 0.14 &    594 & 1.87 & 1.51 &  9.75 \\
\hline
\texttt{Emilia\_923}
 & ABMC              &    64 & 8 & 0.00 & 14,424 & 1.00 & 1.13 &  6.72 \\
 & \texttt{LeidenMD} &    39 & 8 & 0.17 & 42,474 & 1.79 & 1.34 &  4.80 \\
 & \texttt{LeidenCP} &   448 & 9 & 0.16 &  4,138 & 2.01 & 1.37 &  9.64 \\
\hline
\texttt{ecology2}
 & ABMC              &   128 & 5 & 0.00 &  7,813 & 1.00 & 1.25 & 12.7 \\
 & \texttt{LeidenMD} &   144 & 6 & 0.08 & 12,480 & 1.80 & 1.54 &  9.63 \\
 & \texttt{LeidenCP} & 6,366 & 6 & 0.10 &    275 & 1.75 & 1.41 & 11.1 \\
\hline
\texttt{thermal2}
 & ABMC              & 1,024 & 6 & 0.00 &  1,200 & 1.00 & 1.40 & 14.8 \\
 & \texttt{LeidenMD} & 1,099 & 5 & 0.89 & 15,354 & 13.7 & 1.42 &  4.65 \\
 & \texttt{LeidenCP} & 6,404 & 6 & 0.25 &    409 & 2.13 & 1.43 & 12.7 \\
\hline
\texttt{Geo\_1438}
 & ABMC              &   128 & 8 & 0.00 & 11,235 & 1.00 & 1.19 &  9.16 \\
 & \texttt{LeidenMD} &    37 & 6 & 0.13 & 55,883 & 1.44 & 1.11 &  5.78 \\
 & \texttt{LeidenCP} &   721 & 8 & 0.16 &  3,668 & 1.84 & 1.25 &  9.53 \\
\hline
\texttt{Hook\_1498}
 & ABMC              &   512 & 11 & 0.00 &  2,926 & 1.00 & 1.44 &  9.85 \\
 & \texttt{LeidenMD} &    46 &  6 & 0.19 & 51,547 & 1.58 & 1.67 &  5.62 \\
 & \texttt{LeidenCP} &   779 & 10 & 0.16 &  3,576 & 1.86 & 1.68 &  9.63 \\
\hline
\texttt{G3\_circuit}
 & ABMC              &   512 & 10 & 0.00 &  3,097 & 1.00 & 1.48 & 11.0 \\
 & \texttt{LeidenMD} &   116 &  5 & 0.15 & 23,447 & 1.72 & 1.55 & 10.2 \\
 & \texttt{LeidenCP} & 9,777 &  7 & 0.13 &    332 & 2.05 & 1.34 & 10.7 \\
\hline
\end{tabular}
\end{table*}

To quantify this structural non-uniformity, this study introduces the Gini coefficient~\cite{CowellMeasuring2011} as an evaluation index for block-size imbalance.
Given the block sizes denoted by $s_1, s_2, \ldots, s_{n_b}$, the Gini coefficient $G$ is defined as follows:
\begin{equation}
  G = \frac{\sum_{i=1}^{n_b} \sum_{j=1}^{n_b} |s_i - s_j|}{2 (n_b)^2 \left( \frac{1}{n_b} \sum_{i=1}^{n_b} s_i \right)}.
\end{equation}
A value of $G = 0$ represents a perfectly uniform distribution, whereas $G \to 1$ indicates an extremely uneven distribution.

Table~\ref{tab::load_imbalance_stats} summarizes the block- and color-level load-imbalance metrics for the representative blockings used in the performance comparison.
The ratios $B_{\max}/B_{\mathrm{avg}}$ and $C_{\max}/C_{\mathrm{avg}}$ characterize block-level and color-level workload imbalance, respectively.
The $S_{32}$ values are derived from the strong-scaling results in Section~\ref{sec:Scaling}.

For the ABMC method, $B_{\max}/B_{\mathrm{avg}}$ is 1.00 for all matrices because the ABMC method explicitly controls the number of blocks and generates nearly uniform block sizes.
However, the color-wise workload is not perfectly uniform, as indicated by $C_{\max}/C_{\mathrm{avg}}$ values larger than 1.00.
This result shows that even when block sizes are uniform, the workload assigned to each color can still be imbalanced.

For \texttt{LeidenMD}, the degree of block-size imbalance depends strongly on the matrix.
The most severe case is \texttt{thermal2}, for which the Gini coefficient is 0.89 and $B_{\max}/B_{\mathrm{avg}}$ reaches 13.7.
This extreme block-size imbalance explains the poor parallel efficiency of \texttt{LeidenMD} for this matrix, despite its small number of block colors.

In contrast, \texttt{LeidenCP} substantially reduces the block-size imbalance for \texttt{thermal2}, decreasing the Gini coefficient from 0.89 to 0.25 and $B_{\max}/B_{\mathrm{avg}}$ from 13.7 to 2.13.
For the other matrices, \texttt{LeidenCP} maintains $B_{\max}/B_{\mathrm{avg}}$ around two or lower.
These results demonstrate that the CPM function effectively suppresses the formation of excessively large communities and mitigates the sequential processing overhead within individual blocks.

For the Leiden-based methods, the values of $C_{\max}/C_{\mathrm{avg}}$ remain moderate for all matrices, mostly between 1.1 and 1.7.
This indicates that the color-wise workload imbalance is not as severe as the block-size imbalance observed for \texttt{LeidenMD}, especially for \texttt{thermal2}.
For this matrix, these results suggest that oversized blocks are a major contributor to the poor parallel performance of \texttt{LeidenMD}.

\subsection{L2 Cache Efficiency}
Table~\ref{tab::Leiden_ICCG_cache} lists the calculated L2 cache efficiency during the execution of the ICCG method.
The results demonstrate that \texttt{LeidenMD} achieves high L2 hit rates overall; in particular, for \texttt{parabolic\_fem}, \texttt{ecology2}, \texttt{thermal2}, and \texttt{G3\_circuit}, the L2 hit rates reach 0.87 or higher.
These results indicate a cache efficiency comparable to or even higher than that of the ABMC method.
For \texttt{Hook\_1498}, the L2 hit rate improves from 0.69 for ABMC to 0.78 for \texttt{LeidenMD}, indicating that the strongly connected communities extracted as relatively large blocks preserve data locality within each block.

%%%%%%%%%%%%%%%%%%%%%%%%%%%%%%%%%%%%%%%%%%%%%%%%%%%%%%%%%%%%%%
\begin{table}[htb]
\begin{center}
\small
\caption{L2 cache efficiency of Leiden-based ICCG}
\label{tab::Leiden_ICCG_cache}
\begin{tabular}{l|c|c}
\hline
\multicolumn{1}{c|}{ \multirow{2}{*}{Matrix} } & \multicolumn{2}{|c}{\texttt{LeidenMD}} \\ \cline{2-3}
& $R_{\mathrm{hit}}$ & $R_{\mathrm{miss}}$ \rule{0pt}{1.0em} \\
\hline
\texttt{parabolic\_fem}  & 0.87 & 0.060 \\
\texttt{apache2}   &  0.86 & 0.051  \\
\texttt{Emilia\_923}  & 0.80 & 0.039 \\
\texttt{ecology2} & 0.87 & 0.052 \\
\texttt{thermal2} &  0.90 & 0.050 \\
\texttt{Geo\_1438} & 0.78 & 0.040 \\
\texttt{Hook\_1498} & 0.78 & 0.037 \\
\texttt{G3\_circuit} & 0.88 & 0.044 \\
\hline \hline
\multicolumn{1}{c|}{ \multirow{2}{*}{Matrix} } & \multicolumn{2}{|c}{\texttt{LeidenCP}} \\ \cline{2-3}
& $R_{\mathrm{hit}}$ & $R_{\mathrm{miss}}$ \rule{0pt}{1.0em}\\
\hline
\texttt{parabolic\_fem}  & 0.73 & 0.20 \\
\texttt{apache2}   & 0.74  & 0.17  \\
\texttt{Emilia\_923}  & 0.72 & 0.086 \\
\texttt{ecology2} & 0.72 &  0.18 \\
\texttt{thermal2} &  0.83 & 0.10 \\
\texttt{Geo\_1438} & 0.71 & 0.090 \\
\texttt{Hook\_1498} & 0.71 & 0.088 \\
\texttt{G3\_circuit} & 0.74 & 0.16 \\
\hline
\end{tabular}
\end{center}
\end{table}

On the other hand, \texttt{LeidenCP} exhibits lower L2 hit rates and higher L2 miss rates than those of \texttt{LeidenMD} across all matrices.
This degradation is driven by the fact that the CPM function subdivides the blocks, which increases the number of block boundaries and consequently reduces continuous data reuse.
However, despite its lower cache efficiency, \texttt{LeidenCP} achieves shorter execution times than those of \texttt{LeidenMD} for several matrices, as shown in Fig.~\ref{fig::Scatters} and Table~\ref{tab::Leiden_NB_NC}.
This result indicates that ICCG performance is governed not only by cache efficiency, but also by the trade-offs among block size, intra-block sequential workload, number of block colors, parallel load balance, and number of iterations.
%%%%%%%%%%%%%%%%%%%%%%%%%%%%%%%%%%%%%%%%%%%%%%%%%%%%%%%%%%%%%

\subsection{Phase-Time Breakdown}
Table~\ref{tab::phase_timing} presents a breakdown of the ICCG execution time for ABMC, \texttt{LeidenMD}, and \texttt{LeidenCP} using 32 threads.
The execution time is decomposed into the times for forward substitution, backward substitution, and sparse matrix-vector multiplication (SpMV), with the remaining time not shown separately.
The three matrices were selected to represent distinct performance characteristics observed in the experiments:
\texttt{thermal2} exhibits a pronounced slowdown and severe block-size imbalance for \texttt{LeidenMD}, \texttt{parabolic\_fem} shows relatively similar performance among the methods, and \texttt{G3\_circuit} represents a case in which \texttt{LeidenMD} outperforms \texttt{LeidenCP}.

\begin{table*}[htb]
\centering
\caption{Breakdown of ICCG execution time for ABMC, \texttt{LeidenMD}, and \texttt{LeidenCP} using 32 threads.
The values are averages over 10 runs, and the percentages indicate the fraction of the total execution time.}
\label{tab::phase_timing}
\begin{tabular}{l|l|r|rr|rr|rr}
\hline
\multicolumn{1}{c|}{\multirow{2}{*}{Matrix}} &
\multicolumn{1}{c|}{\multirow{2}{*}{Method}} &
\multicolumn{1}{c|}{\multirow{2}{*}{Total [ms]}} &
\multicolumn{2}{c|}{Forward Sub.} &
\multicolumn{2}{c|}{Backward Sub.} &
\multicolumn{2}{c}{SpMV} \\
&
&
&
\multicolumn{1}{c}{Time [ms]} &
\multicolumn{1}{c|}{Ratio [\%]} &
\multicolumn{1}{c}{Time [ms]} &
\multicolumn{1}{c|}{Ratio [\%]} &
\multicolumn{1}{c}{Time [ms]} &
\multicolumn{1}{c}{Ratio [\%]} \\
\hline
\texttt{parabolic\_fem}
 & ABMC
 & 1,790
 & 339 & 18.9
 & 248 & 13.8
 & 176 & 9.9 \\
 & \texttt{LeidenMD}
 & 1,676
 & 297 & 17.7
 & 221 & 13.2
 & 151 & 9.0 \\
 & \texttt{LeidenCP}
 & 1,708
 & 263 & 15.4
 & 236 & 13.8
 & 222 & 13.0 \\
\hline
\texttt{thermal2}
 & ABMC
 & 7,891
 & 1,219 & 15.4
 & 1,239 & 15.7
 & 1,116 & 14.1 \\
 & \texttt{LeidenMD}
 & 23,917
 & 8,911 & 37.3
 & 9,934 & 41.5
 & 975 & 4.1 \\
 & \texttt{LeidenCP}
 & 9,913
 & 1,907 & 19.2
 & 2,117 & 21.3
 & 1,410 & 14.2 \\
\hline
\texttt{G3\_circuit}
 & ABMC
 & 5,501
 & 917 & 16.7
 & 950 & 17.3
 & 622 & 11.3 \\
 & \texttt{LeidenMD}
 & 5,420
 & 1,087 & 20.1
 & 1,231 & 22.7
 & 455 & 8.4 \\
 & \texttt{LeidenCP}
 & 5,880
 & 982 & 16.7
 & 1,024 & 17.4
 & 859 & 14.6 \\
\hline
\end{tabular}
\end{table*}

For \texttt{thermal2}, the forward and backward substitutions account for 37.3\% and 41.5\%, respectively, of the total execution time with \texttt{LeidenMD}, corresponding to 78.8\% in total.
In comparison, the corresponding fractions are 31.1\% for ABMC and 40.5\% for \texttt{LeidenCP}.
The substantially larger substitution times of \texttt{LeidenMD} are consistent with the severe block-size imbalance observed for this matrix, where $B_{\max}/B_{\mathrm{avg}}=13.7$ as shown in Table~\ref{tab::load_imbalance_stats}.
Since the unknowns within each block are processed sequentially, the presence of such large blocks limits the effective parallelism of the forward and backward substitutions.
These measurements therefore indicate that the poor execution performance of \texttt{LeidenMD} for \texttt{thermal2} is strongly associated with the increased cost of the triangular substitution phases.

In contrast, for \texttt{parabolic\_fem}, the substitution-time fractions are relatively similar among the three methods.
For ABMC, \texttt{LeidenMD}, and \texttt{LeidenCP}, the combined forward and backward substitution times account for 32.7\%, 30.9\%, and 29.2\% of the total execution time, respectively.
For \texttt{G3\_circuit}, the corresponding fractions are 34.0\%, 42.8\%, and 34.1\%.
These results indicate that the execution time differences cannot in general be attributed solely to the triangular substitution phases.
Other components, including SpMV and parallel scheduling and synchronization overhead, also contribute to the overall performance.
Thus, the phase-time measurements support the block-imbalance explanation for cases such as \texttt{thermal2}, while also confirming that no single performance factor explains the behavior of all matrices.

%%%%%%%%%%%%%%%%%%%%%%%%%%%%%%%%%%%%%%%%%%%%%%%%
\subsection{Blocking Cost and Repeated-Solve Considerations}
Throughout the performance comparisons in this study, the reported ICCG execution time includes construction of the IC(0) preconditioner and the subsequent iterations, but excludes block generation and block coloring.
The blocking cost is evaluated separately in this subsection.

The ABMC results reported in this study represent a tuned baseline, because they correspond to the best performance among six candidate values of the number of blocks $n_b=32,64,128,256,512,$ and $1024$.
Therefore, the practical cost of obtaining the optimized ABMC configuration includes both the blocking cost and the ICCG execution time for evaluating each of the six candidates.
In contrast, the Leiden-based methods determine the block structure from a single blocking run without explicitly prescribing $n_b$, thereby avoiding this block-count tuning procedure.

For $N$ repeated solves using the same blocking configuration, the total costs of ABMC and a Leiden-based method are therefore expressed as
\begin{equation}
T_{\mathrm{total}}^{\mathrm{ABMC}}(N)
=
T_{\mathrm{tune}}^{\mathrm{ABMC}}
+
N \, T_{\mathrm{solve}}^{\mathrm{ABMC}},
\end{equation}
and
\begin{equation}
T_{\mathrm{total}}^{\mathrm{Leiden}}(N)
=
T_{\mathrm{block}}^{\mathrm{Leiden}}
+
N \, T_{\mathrm{solve}}^{\mathrm{Leiden}},
\end{equation}
respectively.
The total costs are calculated using the measured blocking and ICCG execution times, excluding file input and output operations.
The blocking time includes block coloring, and the ICCG execution time includes construction of the IC(0) preconditioner.
Here, $T_{\mathrm{tune}}^{\mathrm{ABMC}}$ denotes the total tuning cost over the six ABMC candidates, including the blocking time and one ICCG solve for each candidate.
The cost of each ICCG solve is estimated using the mean execution time over 10 runs, rather than the cumulative time of these runs.
$T_{\mathrm{solve}}^{\mathrm{ABMC}}$ denotes the ICCG execution time using the selected ABMC configuration.
For the Leiden-based methods, $T_{\mathrm{block}}^{\mathrm{Leiden}}$ denotes the blocking time of a single Leiden run, and $T_{\mathrm{solve}}^{\mathrm{Leiden}}$ denotes the corresponding ICCG execution time.
The total-cost model includes the matrix-specific block-count tuning cost for ABMC, whereas \texttt{LeidenCP} uses a fixed resolution parameter selected through preliminary experiments.
The cost of these preliminary experiments is not included in the total-cost comparison.
The break-even number of solves $N_{\mathrm{BE}}$ is defined as the smallest integer $N$ for which $T_{\mathrm{total}}^{\mathrm{ABMC}}(N)$ becomes lower than $T_{\mathrm{total}}^{\mathrm{Leiden}}(N)$.
Note that this comparison is asymmetric: the ABMC cost includes matrix-specific tuning of $n_b$, whereas the Leiden cost excludes the one-time selection of $\gamma_{\mathrm{CP}}$.
Excluding this selection cost can overestimate the number of solves for which LeidenCP remains less costly than ABMC.
% The reported $N_{\mathrm{BE}}$ values are therefore upper-side estimates of the advantage of the Leiden-based approach.
We use a single fixed $\gamma_{\mathrm{CP}}$ for all eight matrices (Section~\ref{sec:Gamma_Evaluation} shows robustness over a range from one half to twice the baseline value), so no matrix-specific tuning is required.
%%%%%%%%%%%%%%%%%%%%%%%%%%%%%%%%%%%%%%%%%%%%%%%%%%%

\begin{table*}[htb]
\centering
\caption{ABMC tuning cost over six $n_b$ candidates, Leiden blocking cost, per-solve ICCG execution time, and the resulting break-even number of solves. All execution times are in seconds.}
\label{tab::total_cost_breakeven}
\begin{tabular}{l|rr|rrr|rrr}
\hline
\multicolumn{1}{c|}{\multirow{2}{*}{Matrix}}
& \multicolumn{2}{c|}{ABMC}
& \multicolumn{3}{c|}{\texttt{LeidenMD}}
& \multicolumn{3}{c}{\texttt{LeidenCP}} \\
\cline{2-9}
\rule[-1.5ex]{0pt}{4.0ex}
& $T_{\mathrm{tune}}^{\mathrm{ABMC}}$
& $T_{\mathrm{solve}}^{\mathrm{ABMC}}$
& $T_{\mathrm{block}}^{\mathrm{Leiden}}$
& $T_{\mathrm{solve}}^{\mathrm{Leiden}}$
& $N_{\mathrm{BE}}$
& $T_{\mathrm{block}}^{\mathrm{Leiden}}$
& $T_{\mathrm{solve}}^{\mathrm{Leiden}}$
& $N_{\mathrm{BE}}$ \\
\hline

\texttt{parabolic\_fem}
& 12.5 & 1.79
& 2.23 & 1.68 & --
& 2.32 & 1.71 & -- \\

\texttt{apache2}
& 15.3 & 1.92
& 3.59 & 2.20 & 42
& 3.56 & 2.40 & 25 \\

\texttt{Emilia\_923}
& 133  & 19.2
& 10.0 & 32.9 & 9
& 9.93 & 17.3 & -- \\

\texttt{ecology2}
& 37.2 & 5.09
& 3.83 & 6.18 & 31
& 3.87 & 6.02 & 36 \\

\texttt{thermal2}
& 64.2 & 7.89
& 5.71 & 23.9 & 4
& 5.75 & 9.91 & 29 \\

\texttt{Geo\_1438}
& 51.6 & 6.73
& 16.4 & 9.49 & 13
& 15.9 & 7.25 & 70 \\

\texttt{Hook\_1498}
& 160  & 21.2
& 16.0 & 30.1 & 17
& 15.5 & 22.0 & 192 \\

\texttt{G3\_circuit}
& 52.9 & 5.50
& 7.11 & 5.42 & --
& 7.23 & 5.88 & 121 \\

\hline
\end{tabular}

\vspace{1mm}
\begin{minipage}{0.98\textwidth}
\footnotesize
$N_{\mathrm{BE}}$ denotes the break-even number of solves at which ABMC becomes less costly than the corresponding Leiden-based method.
``--'' indicates that no finite break-even point exists because the per-solve ICCG execution time of ABMC is not lower than that of the corresponding Leiden-based method.
The break-even numbers were calculated from unrounded execution times.
\end{minipage}
\end{table*}

As shown in Table~\ref{tab::total_cost_breakeven}, the upfront cost of the ABMC tuning procedure is higher than that of a single Leiden blocking run for all eight matrices.
When ABMC has a shorter per-solve ICCG execution time, however, this additional tuning cost can be amortized over repeated solves.
For \texttt{LeidenMD}, a finite break-even point is obtained for six of the eight matrices, ranging from 4 solves for \texttt{thermal2} to 42 solves for \texttt{apache2}.
For \texttt{LeidenCP}, a finite break-even point is also obtained for six matrices, but the required number of solves ranges from 25 to 192.
The larger break-even values for several \texttt{LeidenCP} cases reflect its more competitive per-solve ICCG execution time.

No finite break-even point exists when ABMC does not provide a lower per-solve execution time.
This occurs for both Leiden configurations for \texttt{parabolic\_fem}, for \texttt{LeidenCP} for \texttt{Emilia\_923}, and for \texttt{LeidenMD} for \texttt{G3\_circuit}.
These results indicate that Leiden-based blocking can reduce the total cost for a single solve or a small number of solves by avoiding the block-count tuning required for ABMC.
Overall, the relative advantage of the two approaches depends not only on the per-solve ICCG execution time but also on the number of repeated solves over which the blocking or tuning cost is amortized.

%%%%%%%%%%%%%%%%%%%%%%%%%%%%%%%%%%%%%%%%%%%%%%%%%%%%%%%%%%%%%%
\section{Strong-Scaling Evaluation}
\label{sec:Scaling}
\begin{figure*}[htb]
\centering
\includegraphics[width=\linewidth]{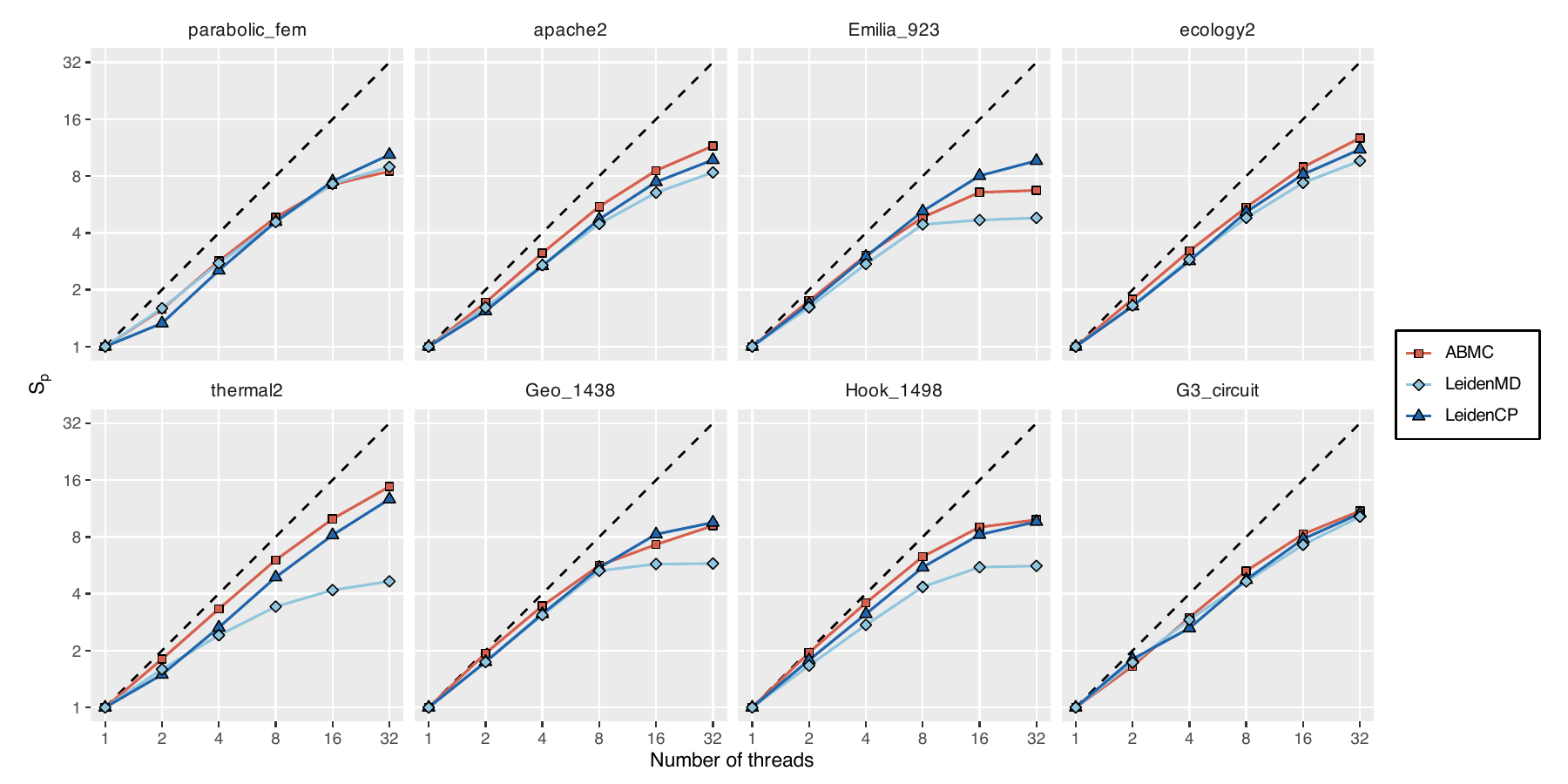}
\caption{Strong-scaling performance of ABMC, \texttt{LeidenMD}, and \texttt{LeidenCP} for the eight test matrices, measured with $p = 1$ to $32$ threads (doubling at each step). The dashed line indicates ideal linear speedup.}
\label{fig::FigStrongScaling}
\end{figure*}

We focus on strong scaling, in which the problem size is fixed and the number of threads is varied, because each test matrix in this study represents a fixed workload.
Fig.~\ref{fig::FigStrongScaling} presents the strong-scaling performance of the ICCG method with parallel preconditioning using ABMC, \texttt{LeidenMD}, and \texttt{LeidenCP}, measured with $p = 1, 2, 4, 8, 16,$ and $32$ OpenMP threads on the 32-core system described in Section~\ref{sec:TestMatrices}.
The speedup is defined as
\begin{equation}
S_p = \frac{T_1}{T_p},
\end{equation}
where $T_p$ denotes the execution time using $p$ threads and $T_1$ denotes the single-thread execution time of each method.
Both the number of threads and $S_p$ are plotted on logarithmic scales.

Because $S_p$ is normalized by the single-thread execution time of each method, it reflects parallel scalability rather than absolute execution performance improvement relative to a common sequential baseline.
Therefore, a smaller value of $S_p$ does not necessarily indicate a longer execution time when the corresponding single-thread execution time is shorter.

\begin{figure*}[htb]
\centering
\includegraphics[width=\linewidth]{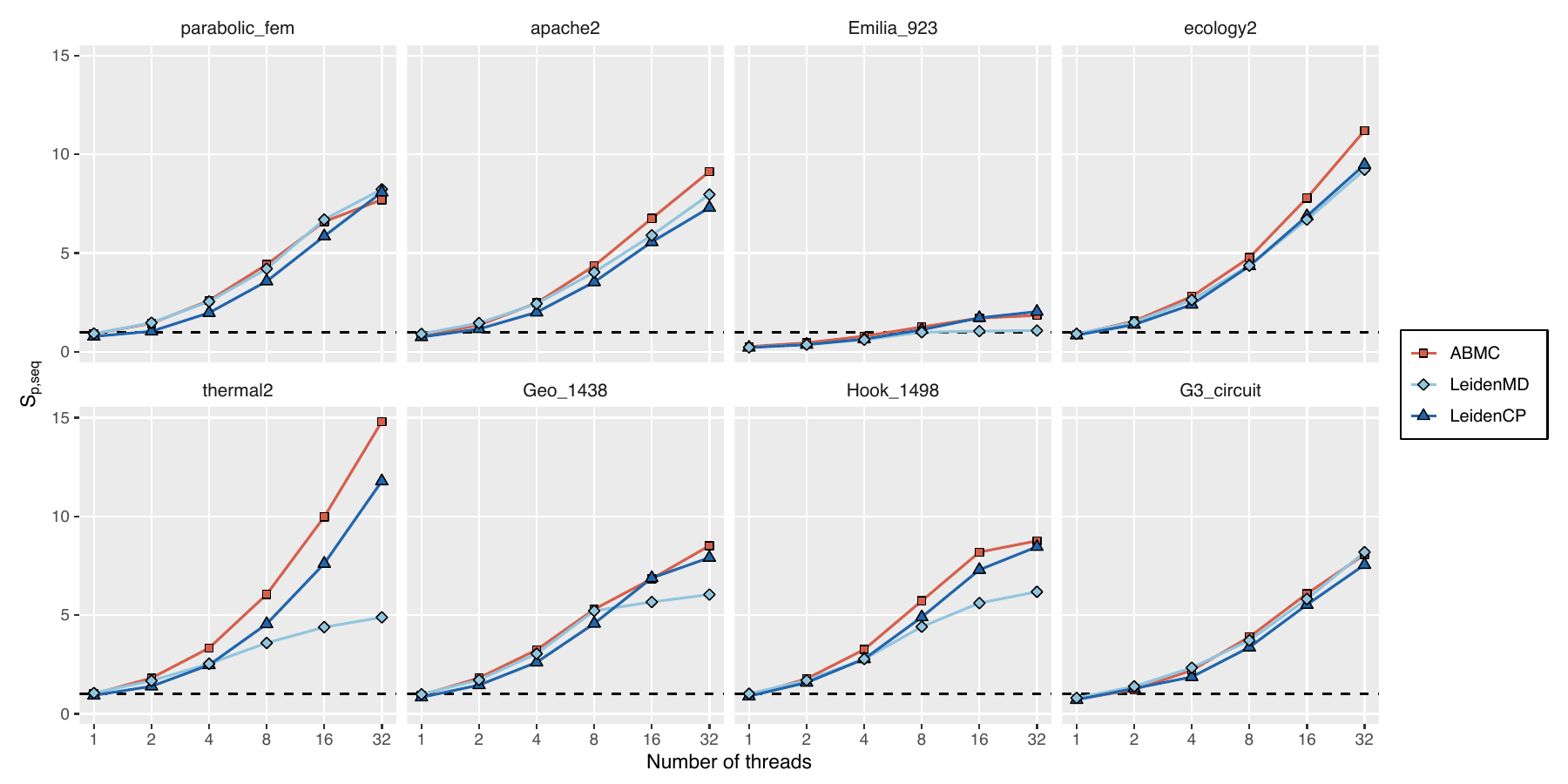}
\caption{Speedup of ABMC, \texttt{LeidenMD}, and \texttt{LeidenCP} relative to the sequential baseline for the eight test matrices.
The dashed line indicates $S_{p,\mathrm{seq}}=1$.}
\label{fig::FigBaseline}
\end{figure*}

To evaluate the absolute performance improvement relative to a common sequential baseline, we additionally define
\begin{equation}
  S_{p,\mathrm{seq}} = \frac{T_{\mathrm{seq}}}{T_p},
\end{equation}
where $T_{\mathrm{seq}}$ is the mean execution time over 10 runs of the sequential IC(0)-ICCG method with the natural ordering (hereafter referred to as the sequential baseline).
The same convergence criterion and right-hand side as in the parallel experiments were used for the sequential baseline.
Fig.~\ref{fig::FigBaseline} shows $S_{p,\mathrm{seq}}$ for the three methods and all eight test matrices.
The horizontal dashed line at $S_{p,\mathrm{seq}}=1$ represents the performance of the sequential baseline; values greater than one therefore indicate a reduction in execution time relative to the sequential baseline.

Table~\ref{tab::baseline_speedup} summarizes the execution time of the sequential baseline, reported as the mean $\pm$ standard deviation over 10 runs, together with the corresponding speedup at 32 threads.
At 32 threads, all three parallel methods achieve $S_{32,\mathrm{seq}}>1$ for all eight matrices.
The improvement, however, varies considerably among the matrices.
For \texttt{Emilia\_923}, the benefit relative to the sequential baseline is limited, with $S_{32,\mathrm{seq}}$ values of 1.84, 1.07, and 2.04 for ABMC, \texttt{LeidenMD}, and \texttt{LeidenCP}, respectively.
For the other matrices, larger improvements are generally obtained, although their magnitude depends on both the matrix and the blocking method.

\begin{table*}[htb]
\centering
\caption{Execution time of the sequential baseline and speedup relative to this baseline at 32 threads.}
\label{tab::baseline_speedup}
\begin{tabular}{l|r|r|r|r}
\hline
\multicolumn{1}{c|}{\multirow{2}{*}{Matrix}}
& \multicolumn{1}{c|}{\multirow{2}{*}{$T_{\mathrm{seq}}$ [s]}}
& \multicolumn{3}{c}{$S_{32,\mathrm{seq}}$} \\
\cline{3-5}
&
& ABMC & \texttt{LeidenMD} & \texttt{LeidenCP} \rule{0pt}{1.0em} \\
\hline
\texttt{parabolic\_fem} &  13.8 $\pm$ 0.036 & 7.71 & 8.23 & 8.08 \\
\texttt{apache2}        &  17.5 $\pm$ 0.038 & 9.13 & 7.97 & 7.30 \\
\texttt{Emilia\_923}    &  35.3 $\pm$ 0.62  & 1.84 & 1.07 & 2.04 \\
\texttt{ecology2}       &  57.0 $\pm$ 0.11  & 11.2 & 9.22 & 9.48 \\
\texttt{thermal2}       &   117 $\pm$ 0.58  & 14.8 & 4.88 & 11.8 \\
\texttt{Geo\_1438}      &  57.3 $\pm$ 0.13  & 8.51 & 6.04 & 7.91 \\
\texttt{Hook\_1498}     &   186 $\pm$ 2.6   & 8.76 & 6.18 & 8.46 \\
\texttt{G3\_circuit}    &  44.4 $\pm$ 0.017 & 8.07 & 8.19 & 7.55 \\
\hline
\end{tabular}
\end{table*}

The strong-scaling behavior can also be interpreted from the block-size statistics in Table~\ref{tab::load_imbalance_stats}.
\texttt{LeidenMD} generally produces a relatively small number of large blocks, as indicated by its large $B_{\max}$ values.
Such coarse-grained blocking can reduce block management and parallelization overhead in low-thread-count execution, but it also limits the number of independent tasks available for parallel execution and can lead to load imbalance as the number of threads increases.
This tendency is particularly evident for \texttt{Emilia\_923}, \texttt{thermal2}, \texttt{Geo\_1438}, and \texttt{Hook\_1498}, for which \texttt{LeidenMD} exhibits relatively poor strong scaling.
For example, \texttt{thermal2} exhibits the largest block-size imbalance among all cases, with $B_{\max}/B_{\mathrm{avg}} = 13.7$ for \texttt{LeidenMD}, and its strong scaling saturates at $S_{32} = 4.65$.

In contrast, \texttt{LeidenCP} generates much finer-grained blocks with substantially smaller $B_{\max}$ values and a much larger number of blocks.
The increased number of parallel tasks generally improves the scalability of \texttt{LeidenCP}, although the finer granularity may also introduce additional scheduling and synchronization overhead.
Nevertheless, for \texttt{apache2}, \texttt{ecology2}, and \texttt{thermal2}, ABMC achieves better strong scaling than \texttt{LeidenCP} despite producing coarser blocks.
This suggests that strong-scaling performance is determined not only by the number and size of blocks but also by the balance among parallel tasks and the overhead associated with block-level synchronization and execution.
Indeed, a correlation analysis between $S_{32}$ and the load-imbalance metrics $G$, $B_{\max}/B_{\mathrm{avg}}$, and $C_{\max}/C_{\mathrm{avg}}$ in Table~\ref{tab::load_imbalance_stats} across the 24 matrix-method combinations revealed only weak-to-moderate linear correlations, with the Pearson correlation coefficient $r$ satisfying $|r| \le 0.47$.
However, the corresponding Spearman rank correlations were not statistically significant ($p>0.05$).
This indicates that no single static block-balance metric is sufficient to fully explain the observed strong-scaling behavior.

%%%%%%%%%%%%%%%%%%%%%%%%%%%%%%%%%%%%%%%%%%%%%%%%%%%%%%%%%%%%%%%%%%%%
\section{Sensitivity Analysis of Resolution Parameters}
\label{sec:Gamma_Evaluation}
In the main experiments, we used $\gamma_{\mathrm{MD}}=1.0$ for \texttt{LeidenMD} and $\gamma_{\mathrm{CP}}=1.0 \times 10^{-3}$ for \texttt{LeidenCP}.
These two parameters belong to different quality functions, and their numerical values are not directly comparable.

For \texttt{LeidenMD}, the standard setting $\gamma_{\mathrm{MD}}=1.0$ often produces coarse partitions with a small number of blocks.
Because such coarse partitions can limit parallelism and increase the sequential processing overhead within large blocks, we also evaluated larger values, namely $\gamma_{\mathrm{MD}}=2.0,4.0,8.0,$ and $16.0$, to examine the effect of finer modularity-based partitions.
For \texttt{LeidenCP}, we varied $\gamma_{\mathrm{CP}}$ from $5.0 \times 10^{-4}$ to $2.0 \times 10^{-3}$, corresponding to one half to twice the baseline value $\gamma_{\mathrm{CP}}=1.0 \times 10^{-3}$.
This range was chosen to evaluate local robustness around the baseline setting, rather than to perform an exhaustive parameter search or to claim the universal optimality of a single parameter value.

\subsection{Effect of $\gamma_{\mathrm{MD}}$}
\label{subsec:GammaMD}
The resolution parameter $\gamma_{\mathrm{MD}}$ in \texttt{LeidenMD} controls the granularity of the partitions obtained via community detection.
In general, a larger value of $\gamma_{\mathrm{MD}}$ yields smaller communities, thereby increasing the total number of blocks $n_b$.

\begin{figure}[htb]
\centering
\includegraphics[width=\linewidth]{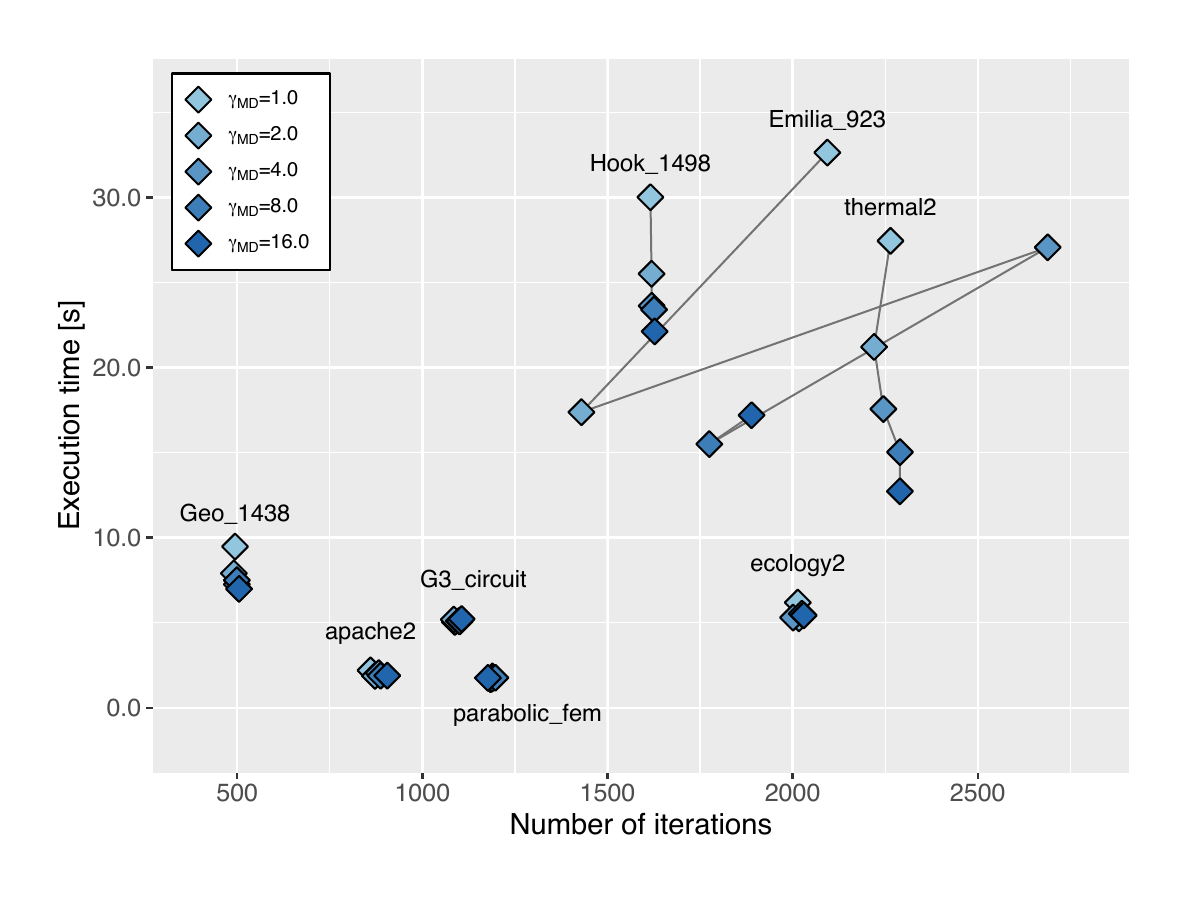}
\caption{Scatter plot of average execution time versus average number of iterations for \texttt{LeidenMD}-based ICCG with various resolution parameter values}
\label{fig::FigLeidenMD_gamma}
\end{figure}

As shown in Table~\ref{tab::BestNB_MC} and \ref{tab::Leiden_NB_NC}, the number of blocks generated by \texttt{LeidenMD} under the default configuration tends to be smaller than that of the optimized ABMC configuration.
Increasing $\gamma_{\mathrm{MD}}$ can potentially generate finer blocks and mitigate the degree of load imbalance.
Fig.~\ref{fig::FigLeidenMD_gamma} illustrates the performance of the ICCG method when the block structures obtained by varying $\gamma_{\mathrm{MD}}$ from 1.0 to 16.0 are applied to the parallel preconditioning.

%%%%%%%%%%%%%%%%%%%%%%%%%%%%%%%%%%%%%%%%%%%%%%%%%%%%%%%%%%%%%%%%%%%%

%%%%%%%%%%%%%%%%%%%%%%%%%%%%%%%%%%%%%%%%%%%%%%%%%%%%%%%%%%%%%%
In terms of execution time, the test matrices exhibit distinct responses to higher $\gamma_{\mathrm{MD}}$ values.
For matrices characterized by severe initial block-size imbalances, such as \texttt{thermal2}, \texttt{Hook\_1498}, and \texttt{Geo\_1438}, increasing $\gamma_{\mathrm{MD}}$ subdivides excessively oversized blocks into finer ones, thereby eliminating the computational bottlenecks associated with prolonged sequential processing within individual blocks.
Consequently, the overall execution time is substantially reduced for \texttt{thermal2} at $\gamma_{\mathrm{MD}}=16.0$.
% markedly reduced, achieving a speedup of approximately 2.16 times for \texttt{thermal2} 
In contrast, for \texttt{parabolic\_fem}, \texttt{apache2}, \texttt{ecology2}, and \texttt{G3\_circuit}, further block refinement yields negligible effects and causes performance saturation.
Because sufficient parallel concurrency is already achieved at the default resolution ($\gamma_{\mathrm{MD}}=1.0$), smaller blocks likely degrade data locality and lower cache efficiency, thereby offsetting the minor gains in parallelism.
Meanwhile, \texttt{Emilia\_923} exhibits a non-monotonic performance trend.
While $\gamma_{\mathrm{MD}}=2.0$ reduces the execution time, higher values increase both the iteration count and the execution time.
These variations demonstrate that finer block partitioning does not straightforwardly improve performance, because the execution time remains highly sensitive to the complex balance among the number of iterations, block size, intra-block sequential workload, and parallel load balancing.

For \texttt{LeidenMD}, these findings confirm that while adjusting $\gamma_{\mathrm{MD}}$ can optimize the execution time for specific matrices, it frequently fails to achieve performance competitive with that of \texttt{LeidenCP}.
This limitation arises from the mathematical definition of modularity, which evaluates the divergence from the expected number of edges based on the global degree distribution.
Consequently, simply increasing $\gamma_{\mathrm{MD}}$ does not necessarily yield block sizes or load balancing tailored for parallel preconditioning in the ICCG method.

Compared with modularity, the CPM is better suited to extracting localized connectivity structures from sparse-matrix graphs because it evaluates community density using a fixed, size-independent threshold.
Moreover, because the constant term $\gamma_{\mathrm{CP}}$ in Eq.~(\ref{eq:cpm}) is applied to every pair of vertices assigned to the same community, its cumulative effect increases with community size, thereby suppressing oversized blocks and reducing block-size imbalance.
These properties contribute to the competitive performance of \texttt{LeidenCP}.
However, because the CPM includes the resolution parameter $\gamma_{\mathrm{CP}}$, its sensitivity to this parameter must also be examined.

%%%%%%%%%%%%%%%%%%%%%%%%%%%%%%%%%%%%%%%%%%%%%%%%%%%%%%%%%%%%%%%
\subsection{Effect of $\gamma_{\mathrm{CP}}$}
\label{subsec:GammaCP}
\begin{figure}[htb]
\centering
\includegraphics[width=\linewidth]{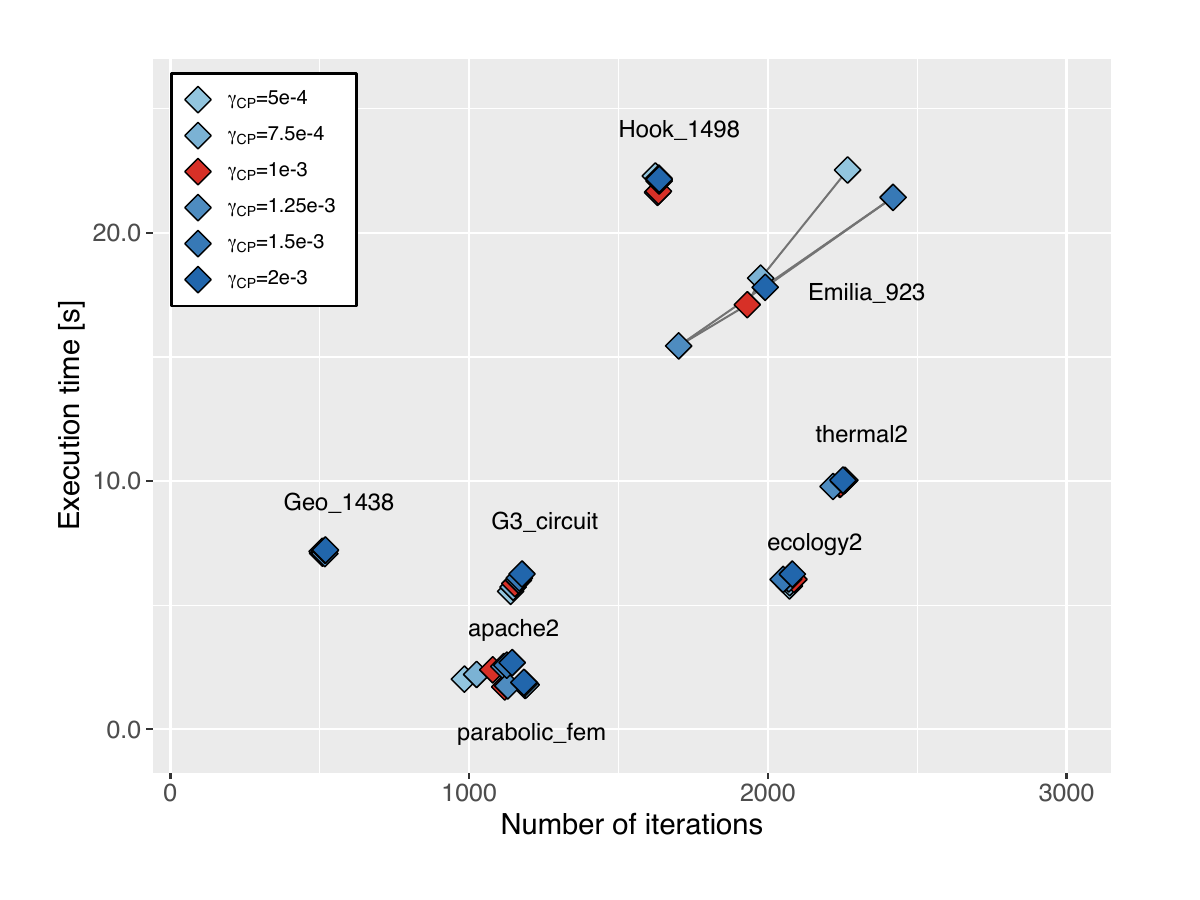}
\caption{Scatter plot of average execution time versus average number of iterations for \texttt{LeidenCP}-based ICCG with various resolution parameter values}
\label{fig::FigLeidenCP_gamma}
\end{figure}

For \texttt{LeidenCP}, we evaluate how the CPM resolution parameter $\gamma_{\mathrm{CP}}$ affects the block structure and ICCG performance.
Table~\ref{tab::gamma_cp_block_status} summarizes the corresponding block statistics.
Fig.~\ref{fig::FigLeidenCP_gamma} shows the relationship between the average execution time and the average number of iterations for various values of $\gamma_{\mathrm{CP}}$.

\begin{table*}[htb]
\caption{Sensitivity of \texttt{LeidenCP} block statistics to $\gamma_{\mathrm{CP}}$; each entry is shown in the form $n_b / n_c / G$, where $n_b$ is the number of blocks, $n_c$ is the number of block colors, and $G$ is the Gini coefficient of block sizes}
\label{tab::gamma_cp_block_status}
\centering
\small
\setlength{\tabcolsep}{2pt}
\begin{tabular}{l | *{5}{r@{ / }c@{ / }l |} r@{ / }c@{ / }l}
\hline
\multicolumn{1}{c|}{ \multirow{2}{*}{Matrix} }
& \multicolumn{18}{c}{$\gamma_{\mathrm{CP}}$} \\
\cline{2-19}
\rule{0pt}{2.4ex}
& \multicolumn{3}{c|}{$5.0 \times 10^{-4}$}
& \multicolumn{3}{c|}{$7.5 \times 10^{-4}$}
& \multicolumn{3}{c|}{$1.0 \times 10^{-3}$}
& \multicolumn{3}{c|}{$1.25 \times 10^{-3}$}
& \multicolumn{3}{c|}{$1.5 \times 10^{-3}$}
& \multicolumn{3}{c}{$2.0 \times 10^{-3}$} \\ 
\hline
\texttt{parabolic\_fem}
& 1,418 & 6 & 0.04
& 1,868 & 6 & 0.04
& 2,277 & 6 & 0.05
& 2,672 & 6 & 0.05
& 2,995 & 6 & 0.05
& 3,687 & 6 & 0.05 \\
\texttt{apache2}
& 1,185 & 6 & 0.12
& 1,718 & 7 & 0.14
& 2,253 & 8 & 0.14
& 2,809 & 8 & 0.14
& 3,311 & 8 & 0.14
& 4,164 & 8 & 0.14 \\
\texttt{Emilia\_923}
& 253 & 9 & 0.17
& 356 & 10 & 0.17
& 448 & 9 & 0.16
& 532 & 9 & 0.15
& 627 & 10 & 0.16
& 794 & 10 & 0.16 \\
\texttt{ecology2}
& 3,941 & 6 & 0.09
& 5,227 & 6 & 0.10
& 6,366 & 6 & 0.10
& 7,370 & 6 & 0.10
& 8,371 & 6 & 0.10
& 10,134 & 6 & 0.10 \\
\texttt{thermal2}
& 4,368 & 6 & 0.32
& 5,448 & 6 & 0.28
& 6,404 & 6 & 0.25
& 7,335 & 6 & 0.24
& 8,165 & 6 & 0.22
& 9,749 & 6 & 0.20 \\
\texttt{Geo\_1438}
& 405 & 9 & 0.16
& 581 & 9 & 0.16
& 721 & 8 & 0.16
& 853 & 9 & 0.15
& 980 & 9 & 0.16
& 1,224 & 9 & 0.16 \\
\texttt{Hook\_1498}
& 440   & 9  & 0.15
& 622   & 9  & 0.15
& 779   & 10 & 0.16
& 921   & 9  & 0.16
& 1,077 & 9  & 0.16
& 1,354 & 9  & 0.16 \\
\texttt{G3\_circuit}
& 6,065 & 7 & 0.13
& 8,020 & 7 & 0.13
& 9,777 & 7 & 0.13
& 11,363 & 7 & 0.13
& 12,828 & 7 & 0.13
& 15,681 & 7 & 0.13 \\
\hline
\end{tabular}
\end{table*}

From Table~\ref{tab::gamma_cp_block_status}, as $\gamma_{\mathrm{CP}}$ increases, the number of blocks $n_b$ generally increases for all test matrices, indicating that larger $\gamma_{\mathrm{CP}}$ values yield finer partitions.
However, the number of block colors $n_c$ remains within a relatively narrow range.
For most matrices, $n_c$ stays between 6 and 10 throughout the tested range of $\gamma_{\mathrm{CP}}$.
This indicates that increasing $\gamma_{\mathrm{CP}}$ refines the block partition while keeping the number of block colors, and hence the synchronization overhead, relatively small.

The Gini coefficient of block sizes also remains moderate over the tested range of $\gamma_{\mathrm{CP}}$.
For example, \texttt{parabolic\_fem}, \texttt{apache2}, \texttt{ecology2}, \texttt{Emilia\_923}, \texttt{Geo\_1438}, \texttt{Hook\_1498}, and \texttt{G3\_circuit} maintain Gini coefficients below approximately 0.17.
Although \texttt{thermal2} exhibits greater block-size imbalance than the other matrices, its Gini coefficient decreases from 0.32 to 0.20 as $\gamma_{\mathrm{CP}}$ increases from $5.0 \times 10^{-4}$ to $2.0 \times 10^{-3}$.
These results indicate that the CPM, through its resolution parameter $\gamma_{\mathrm{CP}}$, effectively suppresses the formation of excessively large blocks over the tested range of $\gamma_{\mathrm{CP}}$ values.

The ICCG performance in Fig.~\ref{fig::FigLeidenCP_gamma} further shows that the performance of \texttt{LeidenCP} is not determined solely by the number of generated blocks.
Finer partitions can improve load balance by reducing oversized blocks, but excessive refinement may also reduce intra-block data locality and change the quality of the preconditioner.
Consequently, the execution time and the number of iterations do not necessarily vary monotonically with $\gamma_{\mathrm{CP}}$.
Nevertheless, the results show that the baseline value $\gamma_{\mathrm{CP}}=1.0 \times 10^{-3}$ is not an isolated or exceptional setting.
Rather, it lies within a practical range for which \texttt{LeidenCP} produces balanced block structures and competitive ICCG performance.
These observations support the robustness of the CPM-based blocking approach with respect to moderate changes in $\gamma_{\mathrm{CP}}$.

%%%%%%%%%%%%%%%%%%%%%%%%%%%%%%%%%%%%%%%%%%%%%%%%%%%%%%%%%%%%%%%
\section{Conclusion and Future Work}
\label{sec:conclusion}
In this study, we proposed a Leiden-based blocking framework for parallel preconditioning in the ICCG method and evaluated its performance against the ABMC method.
Unlike ABMC, which requires the number of blocks to be specified in advance, the proposed framework determines block structures by optimizing a graph-partitioning quality function without prescribing the number of blocks, although a resolution parameter still has to be set.
Compared with the modularity-based variant (\texttt{LeidenMD}), \texttt{LeidenCP} suppresses the formation of excessively large blocks and provides more balanced block structures for parallel ICCG.

The ICCG execution time of \texttt{LeidenCP} was within 10\% of that of the finely tuned ABMC method for five of the eight test matrices and was shorter for two, despite an average L2 hit rate 7.6 percentage points lower than that of ABMC.
The total-cost evaluation further showed that Leiden-based blocking has a lower upfront cost than tuned ABMC because it avoids block-count tuning, whereas ABMC can amortize its higher tuning cost over repeated solves when its per-solve ICCG execution time is lower.
This evaluation excludes the cost of the preliminary experiments used to select the \texttt{LeidenCP} resolution parameter.
Thus, within this cost model, the practical choice between ABMC and Leiden-based blocking depends on both per-solve performance and the number of solves.

%%% Added 20261103
The present evaluation covers eight symmetric positive definite  matrices on a single 32-core shared-memory system; validation on larger systems, including weak scaling of graph construction, blocking, and coloring, and execution on higher-core-count and multi-socket non-uniform memory access platforms, remains for future work.
Methodological extensions include other quality functions, parallel block generation~\cite{ZengParallel2015}, locality-oriented reordering such as Rabbit Order~\cite{AraiRabbit2016}, and incomplete LU preconditioning.
Preconditioning quality was assessed empirically through ICCG iteration counts; future numerical analysis will examine preconditioned spectra and matrix characteristics associated with IC(0) breakdown.
The absence of breakdown in the present tests does not imply a general guarantee.
%%%%%%%%%%%%%%%%%%%%%%%%%%%%%%%%%%%%%%%%%%%%%%%%%%%%%%%%%%%%%%%

\bibliographystyle{IEEEtran}
\bibliography{IEEEabrv,stomo}

%\begin{IEEEbiography}{Tomohiro Suzuki} received bachelor's and master's degrees from the University of Yamanashi, Japan, in 1989 and 1991, respectively, and a Ph.D. degree from Nagoya University in 2003.
%  He is currently a Professor at the University of Yamanashi.
%  He received the 2000 Japan Journal of Industrial and Applied Mathematics Best Paper Award.
%  He is a member of the ACM, the Japan Society for Industrial and Applied Mathematics (JSIAM) and the Information Processing Society of Japan (IPSJ).
%\end{IEEEbiography}

% \EOD

\end{document}